\documentclass[aps,prl,reprint,superscriptaddress,nobibnotes]{revtex4-2}

\usepackage{mathptmx}
\usepackage{graphicx}
\usepackage{dcolumn}
\usepackage{bm}
\usepackage{amsmath,amssymb,amsfonts}
\usepackage{hyperref}
\hypersetup{hidelinks}
\usepackage{xcolor}
\usepackage[normalem]{ulem}
\usepackage{etoolbox}

\newcommand{\Dpt}{D_{\mathrm{pt}}}           % perturbative-control dimension
\newcommand{\epspt}{\epsilon_{\mathrm{pt}}}
\newcommand{\Nfcr}{N_f^{\mathrm{cr}}}
\newcommand{\SU}{\mathrm{SU}}

\newcommand{\MSbar}{\overline{\mathrm{MS}}}

\newcommand{\redstrike}{\bgroup\markoverwith{\textcolor{red}{\rule[0.5ex]{2pt}{0.4pt}}}\ULon}

\newcommand{\add}[1]{\textcolor{blue}{#1}}

\definecolor{vpurple}{RGB}{150,0,200}

\newcommand{\xir}{\xi}
\newcommand{\dd}{\mathrm{d}}

\newcommand{\EMpanel}[3]{%
  \begin{minipage}[t]{0.315\textwidth}
    \centering
    \includegraphics[
      width=\linewidth,
      height=#3\textheight,
      keepaspectratio
    ]{#1}\par
    \vspace{1pt}
    {\footnotesize #2\par}
  \end{minipage}%
}

\makeatletter
\long\def\@author@join@#1#2#3{%
  \def\@author{{#1}{\relax#2\rtxnote{#3}}}}
\protected\long\def\rtxnote#1{%
  \rtxsep\frontmatter@footnote{#1}%
  \def\rtxsep{\textsuperscript{,\,}}}
\patchcmd{\doauthor}{\frontmatter@footnote{#2}}
  {{\let\rtxsep\@empty#2}}{}
  {\PackageError{authornotes}{REVTeX patch failed}{}}
\makeatother

\newcommand{\EqualContributionText}{%
Authors are listed alphabetically by surname, following common
practice in high-energy physics. Jie Liu and Yu-Bo Zhao contributed
equally to this work and share first authorship.%
}

\begin{document}

\title{Asymptotic Safety in Gauge Theories beyond Four Dimensions: \\
Perturbative Tests and Fixed-Point Mergers}

\author{Aldo~Deandrea}
\email{deandrea@ip2i.in2p3.fr}
\affiliation{Universit\'e Lyon 1, CNRS, IP2I, UMR 5822, Villeurbanne, France.}\affiliation{Department of Physics, University of Johannesburg, PO Box 524, Auckland Park 2006, South Africa.}

\author{Jie Liu}
\email{liujie1021@uestc.edu.cn}
\thanks{\EqualContributionText}
\affiliation{School of Physics, The University of Electronic Science and Technology of China,\\
No.~2006, Xiyuan Ave, West Hi-Tech Zone, Chengdu, China}

\author{Roman Pasechnik}
\email{roman.pasechnik@fysik.lu.se}
\affiliation{
 Department of Physics, Lund University
 SE-223 62 Lund, Sweden
}

\author{Zhi-Wei Wang}
\email{Corresponding Author: zhiwei.wang@uestc.edu.cn}
\affiliation{School of Physics, The University of Electronic Science and Technology of China,\\
No.~2006, Xiyuan Ave, West Hi-Tech Zone, Chengdu, China}

\author{Yu-Bo Zhao}
\email{yubo.zhao@std.uestc.edu.cn}
\thanks{\EqualContributionText}
\affiliation{School of Physics, The University of Electronic Science and Technology of China,\\
No.~2006, Xiyuan Ave, West Hi-Tech Zone, Chengdu, China}

\begin{abstract}
Non-Abelian gauge theories above four dimensions may admit an interacting ultraviolet
fixed point through the competition between canonical scaling and gauge-field
antiscreening.
Using the four-loop $\MSbar$ beta function continued to $D=4+2\epsilon$, we determine
the largest dimension $\Dpt=4+2\epspt$ below which successive loop corrections to
the fixed point remain hierarchically ordered under a specified ratio criterion, and map $\Dpt$ over color and flavor numbers for fermions in the fundamental representation of $SU(N)$. At $D=5$, we construct rational anomalous dimensions with a fixed,
pole-regularized denominator inspired by the functional renormalization group (fRG).
Matching to the $\MSbar$ series through one, two, and three loops produces progressively smaller shifts in the critical flavor number and coupling at the ultraviolet--infrared merger (the point where the infrared and ultraviolet fixed points coincide), yielding $N_f^{\mathrm{cr}}\simeq 2.09 \, N$ at large $N$ in the three-loop construction.
This low-order stabilization, together with qualitative evidence from fRG flows
including higher gauge operators, suggests a potential merger scenario that bounds the candidate asymptotically safe region in flavor number. The resulting color--flavor map provides guidance for five-dimensional model building by displaying both the baseline perturbative-control boundary and the candidate three-loop merger boundary.
Four-loop matching, however, retains an interacting ultraviolet fixed point beyond
the three-loop merger boundary, with no merger found for integer $2\le N\le15$
and $0\le N_f/N\le11/2$.
The sensitivity of the fixed-point structure to higher-order terms
leaves both scenarios open, highlighting fixed-point annihilation
as a potential constraint on five-dimensional asymptotic safety.
\end{abstract}

\maketitle

\section{Introduction}

Non-Abelian gauge theories in spacetime dimensions larger than four are nonrenormalizable by conventional power counting, but may admit an interacting ultraviolet (UV) fixed point.
Such a fixed point can arise from the competition between the positive canonical contribution to the beta function of the dimensionless gauge coupling and gauge-field antiscreening.
This possibility has long been studied in the context of extra-dimensional gauge dynamics and asymptotic safety~\cite{weinberg1979ultraviolet,peskinCRITICALPOINTBEHAVIOR1980,kazakovUltravioletFixedPoints2002,dienesGUTPrecursorsNonTrivial2003,giesRenormalizabilityGaugeTheories2003,morrisRenormalizableExtradimensionalModels2005,DeCesare:2021,Codello:2016}.
Such fixed points also motivate phenomenological model building in extra dimensions, including asymptotic grand unification~\cite{abdalgabarUnificationGaugeYukawa2018,cacciapagliaMinimalSU5Asymptotic2021,cacciapagliaAsymptoticUnification2024,fangAsymptoticGrandUnification2025,Cacciapaglia:2026fqg}. The search for a continuum limit has also motivated lattice studies: for five-dimensional $\SU(2)$ Yang--Mills theory, a continuous transition was excluded in the sampled parameter range of the Wilson action extended by an adjoint plaquette~\cite{florioSearchingContinuousPhase2021}. Assessing the reliability of the underlying fixed points and their dependence on spacetime dimension and matter content is therefore important both for ultraviolet completion and for phenomenological model building.

Two properties distinguish $D>4$ from the familiar four-dimensional situation. First,
the fixed point is located at finite coupling, so any statement about it comes with a
question of convergence. Second, because the beta function of the dimensionless
coupling starts with the canonical term $\tfrac12(D-4)a$, even its ``one-loop''
coefficient is no longer universal: under an analytic redefinition $a\to a+c\,a^2$ it
shifts by $c(D-4)/2$. Different regularizations can therefore yield different one-loop beta-function coefficients for the same five-dimensional theory, so quantitative estimates should be accompanied by an assessment of their renormalization-scheme dependence.

A controlled way to address the first question is to continue the theory away from
four dimensions, $D=4+2\epsilon$, and determine the fixed point order by order in
$\epsilon$. The availability of four-loop $\MSbar$ beta functions for non-Abelian gauge
theories with fermions \cite{herzogFiveloopBetaFunction2017,baikovFiveLoopRunningQCD2017}
makes it possible to track the loop-order dependence of the fixed-point coupling and to
diagnose the loss of perturbative control. We use this information to define a
\emph{perturbative-control dimension} $\Dpt=4+2\epspt$, above which the hierarchy of
successive loop corrections to the UV fixed point is no longer maintained. $\Dpt$ is a
property of the perturbative series, not of the theory; it should not be confused with the dynamical critical dimension of Ref.~\cite{giesRenormalizabilityGaugeTheories2003}, above which the fixed point itself
ceases to exist.

Finite-order perturbative results do not by themselves establish the global
structure of the gauge-coupling flow in five dimensions. A candidate scenario is the coexistence of an interacting UV fixed point
with an infrared (IR) partner~\cite{giesRenormalizabilityGaugeTheories2003}. Fermionic screening can then
drive the merger and annihilation of the two fixed points, providing a candidate upper flavor boundary $\Nfcr$ for the conformal window
\cite{Kaplan.DConformality-Lost2009,pastor-gutierrezPhaseStructureExtradimensional2024,goertzGaugeFermionCartographyConfinement2024}.
The origin of this IR partner need not coincide with the fermion-induced
Banks--Zaks mechanism in four dimensions~\cite{caswell1974asymptotic,banks1982phase},
since a similar fixed-point structure has been proposed for pure
Yang--Mills theory in extra dimensions~\cite{giesRenormalizabilityGaugeTheories2003}. To investigate this possibility, we use a rational ansatz for the gauge
anomalous dimension inspired by the functional renormalization group
(fRG)~\cite{wetterichExactEvolutionEquation1993,Morris.TThe-Exact-Renormalisation-Group1994,bergesNonPerturbativeRenormalizationFlow2002,pawlowskiAspectsFunctionalRenormalisation2007}. Its weak-coupling expansion is matched to the $\MSbar$ series, while its denominator incorporates a fRG-inspired resummation
structure~\cite{giesRunningCouplingYangMills2002,giesIntroductionFunctionalRG2012,pastor-gutierrezUVCompletionExtradimensional2023}. This construction allows us to explore the candidate merger and assess its dependence on the perturbative matching order.

The main results of this Letter are twofold. First, we determine the UV fixed point at successive loop orders using the
four-loop $\MSbar$ beta function continued to $D=4+2\epsilon$, and map the
perturbative-control dimension $\Dpt$ across the $(N,N_f)$ plane for $SU(N)$
theories with fundamental Dirac fermions. Second, at $D=5$, we construct rational beta functions with a common,
pole-regularized denominator inspired by fRG, matching their weak-coupling
expansions to the $\MSbar$ series at successive orders. 
Matching through one, two, and three loops produces progressively smaller
shifts in the critical flavor number and coupling at the UV--IR merger,
yielding $\Nfcr\simeq2.09N$ at large $N$ in the three-loop construction,
about $10\%$ below the unresummed three-loop estimate. 
Combining the perturbative-control map with the candidate merger boundary
provides a color--flavor phase diagram that guides five-dimensional model
building.
Within the gauge-coupling projection, for $2\le N\le15$, four-loop matching with the same denominator instead retains an interacting UV fixed point beyond the
three-loop merger boundary, with no merger over the color and flavor
ranges studied, allowing a larger candidate domain of asymptotic safety beyond the merger bound. Both scenarios remain open within the present analysis, making UV--IR annihilation a potentially important limitation on five-dimensional ultraviolet completion.

\section{UV fixed point from the $\epsilon$ expansion}

We consider a $\SU(N)$ gauge theory in $D>4$ dimensions. The beta function of the dimensionless renormalized coupling $a=g^2/16\pi^2$ is defined as
\begin{equation}
	\beta^{(D)}(a)\equiv \frac{1}{2}\frac{\dd a}{\dd \ln \mu},
\end{equation}
where, in the $\MSbar$ scheme continued to $D$ dimensions, the canonical scaling
contribution separates from the four-dimensional beta function,
\begin{eqnarray}
	\beta^{(D)}(a)&=&\frac{1}{2}(D-4)\,a+\beta^{(D=4)}(a),
	\nonumber \\
	\mathrm{with} && \beta^{(D=4)}(a)=-\sum_{n=0}^\infty\beta_n a^{n+2}.
	\label{eq:beta_split}
\end{eqnarray}
Equation~\eqref{eq:beta_split} represents the one-coupling $\MSbar$ projection of the effective theory onto the $F^2$ sector, rather than the complete RG flow of a general $D$-dimensional gauge theory. The local operator expansion also contains interactions such as
$(DF)^2$ and $F^3$, written schematically, which have canonical
dimension six in four dimensions.
Their canonical irrelevance at the four-dimensional Gaussian fixed
point motivates this truncation for sufficiently small $\epsilon$
\cite{morrisRenormalizableExtradimensionalModels2005}. At an interacting fixed point, however, their relevance also depends on anomalous dimensions and operator mixing; canonical counting alone does not establish the validity of this truncation throughout $4<D<6$. The loop-order hierarchy studied below tests perturbative stability
within this projection and does not independently assess the effects
of omitted operators.

\begin{figure}[t]
    \centering
    \includegraphics[width=0.45\textwidth]{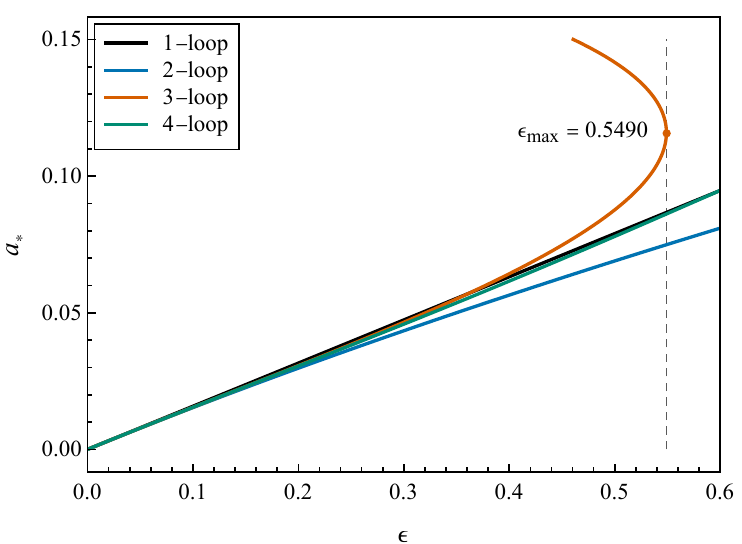}
    \caption{Fixed-point couplings as functions of the extra-dimensional parameter
$\epsilon$ at successive loop orders for an $SU(3)$ gauge theory with $N_f=7$ fundamental Dirac fermions. The three-loop curve includes both the lower-coupling UV branch and
the upper-coupling IR branch, which merge at $\epsilon_{\mathrm{max}}\simeq0.549$ and $a_*\simeq0.116$.
The one-, two-, and four-loop curves show the UV fixed-point branches.}
    \label{fig:fixedpoint_values}
\end{figure}

The loop coefficients $\beta_n$ depend on the gauge group, the number of fermion flavors and their representations, and are considered here up to four loops
\cite{herzogFiveloopBetaFunction2017,baikovFiveLoopRunningQCD2017}. For $D=4+2\epsilon$
the non-Gaussian UV fixed point $a_*$ is determined by
\begin{equation}
	\beta^{(4+2\epsilon)}(a_*)=\epsilon a_*-\sum_{n=0}^3\beta_n (a_*)^{n+2}=0.
	\label{eq:beta_epsilon}
\end{equation}
The $L$-loop fixed-point values $a_*^{(L)}(\epsilon)$ are obtained from the truncated
equations
\begin{equation}
	\epsilon=\sum_{n=0}^{L-1}\beta_n\bigl(a_*^{(L)}\bigr)^{n+1},\qquad L=1,\dots,4,
	\label{eq:truncated_equations}
\end{equation}
For $\beta_0>0$, we select the real positive branch emerging from the
Gaussian fixed point $a=0$, characterized by
$a_*^{(L)}(\epsilon)=\epsilon/\beta_0+O(\epsilon^2)$ as
$\epsilon\to0^+$. We then follow this branch as $\epsilon$ increases,
 as long as it remains real and positive. As a representative benchmark, Fig.~\ref{fig:fixedpoint_values} shows
$a_*^{(L)}(\epsilon)$ for $N=3$ and $N_f=7$ fundamental Dirac fermions, a flavor number chosen near the estimated three-loop conformal-window boundary discussed below. Over the plotted range, the two- and four-loop UV values track the one-loop line $a_*^{(1)}=\epsilon/\beta_0$, whereas the three-loop curve folds back.
For this benchmark, $\beta_2\simeq-233.87$, and the right-hand side
of Eq.~\eqref{eq:truncated_equations} at $L=3$, viewed as a function
of $a>0$, reaches its maximum $\epsilon_{\max}\simeq0.549$
at $a\simeq0.116$. At this turning point, the lower-coupling UV branch meets the
upper-coupling IR branch; no positive three-loop fixed point remains
for $\epsilon>\epsilon_{\max}$. With $\beta_3\simeq982.56$, the right-hand side of Eq.~\eqref{eq:truncated_equations} at $L=4$ is instead
monotonically increasing for $a>0$, yielding a unique positive UV
fixed point for every $\epsilon>0$ within this truncation.

\section{Perturbative-control dimension}

The fixed-point solutions are perturbatively meaningful only as long as the expansion in
$\epsilon$ remains under control. As $\epsilon$ increases, $a_*$ moves toward larger
values and higher-loop corrections become comparable to lower-order ones. To quantify
this behavior, we decompose the four-loop fixed-point value into successive loop increments,
\begin{equation}
	c_1(\epsilon)\equiv a_*^{(1)}(\epsilon),\quad
	c_L(\epsilon)\equiv a_*^{(L)}(\epsilon)-a_*^{(L-1)}(\epsilon),\quad L=2,3,4 .
	\label{eq:increments}
\end{equation}
We test the hierarchy $|c_1|>|c_2|>|c_3|>|c_4|$ using the ratios $R_L=|c_L/c_{L-1}|$ for $L=2,3,4$. Following the real weak-coupling branches from $\epsilon\to0^+$,
we require all three ratios to remain below a threshold $\theta$,
with $\theta=1/2$ as our baseline. For each $L$, let $\epsilon_L$ denote the first value of $\epsilon$ at which $R_L$ reaches $\theta$. The earliest of these crossings defines
\begin{equation}
	\epspt \equiv \min_{L}\;\epsilon_L\,,\qquad
	R_L(\epsilon_L)\equiv\Bigl|\frac{c_L(\epsilon_L)}{c_{L-1}(\epsilon_L)}\Bigr|=\theta\,.
	\label{eq:criterion}
\end{equation}
The corresponding perturbative-control dimension is
\begin{equation}
	\Dpt\left(N,N_f\right) = 4 + 2\epspt\,.
\end{equation}
We retain the first threshold crossing as the definition of $\epspt$,
even if all ratios fall below the threshold again at larger $\epsilon$. It is a scheme- and threshold-dependent diagnostic, not an error bound or a necessary condition for small absolute corrections. A small denominator $c_{L-1}$ can inflate the ratios $R_L$ even when the absolute correction is small while a vanishing denominator renders it undefined.
Alternatively, a fixed-point branch may end (UV-IR merger scenario, discussed later)
before any ratio reaches $\theta$, we record this case separately (Appendix~\ref{app:hierarchy}).

\begin{figure}[t]
    \centering
    \includegraphics[width=0.45\textwidth]{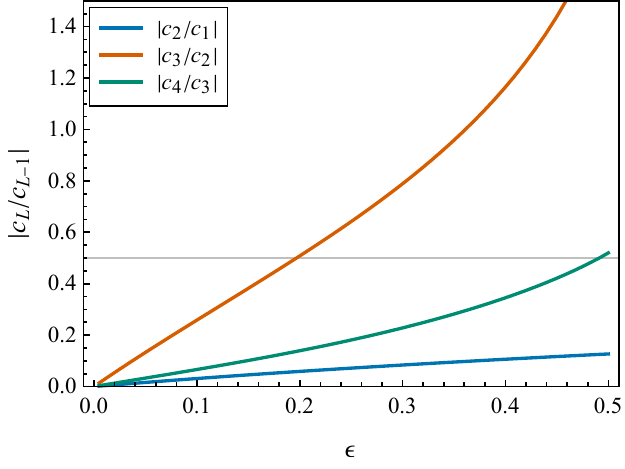}
    \caption{Ratios of successive loop increments, $|c_L/c_{L-1}|$, as functions of
    $\epsilon$ for $\SU(3)$ with $N_f=7$. Indication that the perturbative control is lost occurs when the first
    ratio reaches the threshold $\theta$ (grey line, $\theta=1/2$).}
    \label{fig:ratios}
\end{figure}

Fig.~\ref{fig:ratios} shows the increment ratios for the benchmark $\SU(3)$ theory with $N_f=7$. The baseline threshold is first reached by $R_3$ at $\epsilon\simeq0.20$, giving $\Dpt(3,7)\simeq4.4$. Over the plotted range, $R_2$ remains below $0.13$, while $R_4$
reaches $1/2$ only at $\epsilon\simeq0.49$. This early crossing reflects the enhanced three-loop correction associated with the upward bending of the branch in Fig.~\ref{fig:fixedpoint_values}. For comparison, at $N_f=2$ the baseline criterion remains satisfied up to $\epsilon=0.5$, although $R_4\simeq0.495$ is already close
to the threshold. The first crossing is then set by the four-loop increment:
$R_4$ reaches $1/2$ at $\epsilon\simeq0.516$, yielding $\Dpt(3,2)\simeq5.03$. The sensitivity of the resulting classification to the choice of threshold, $\theta\in\{0.5,0.8,1\}$, is shown in Fig.~\ref{fig:threshold_sensitivity} of the Appendix.

\begin{figure}[t]
    \centering
    \includegraphics[width=\linewidth]{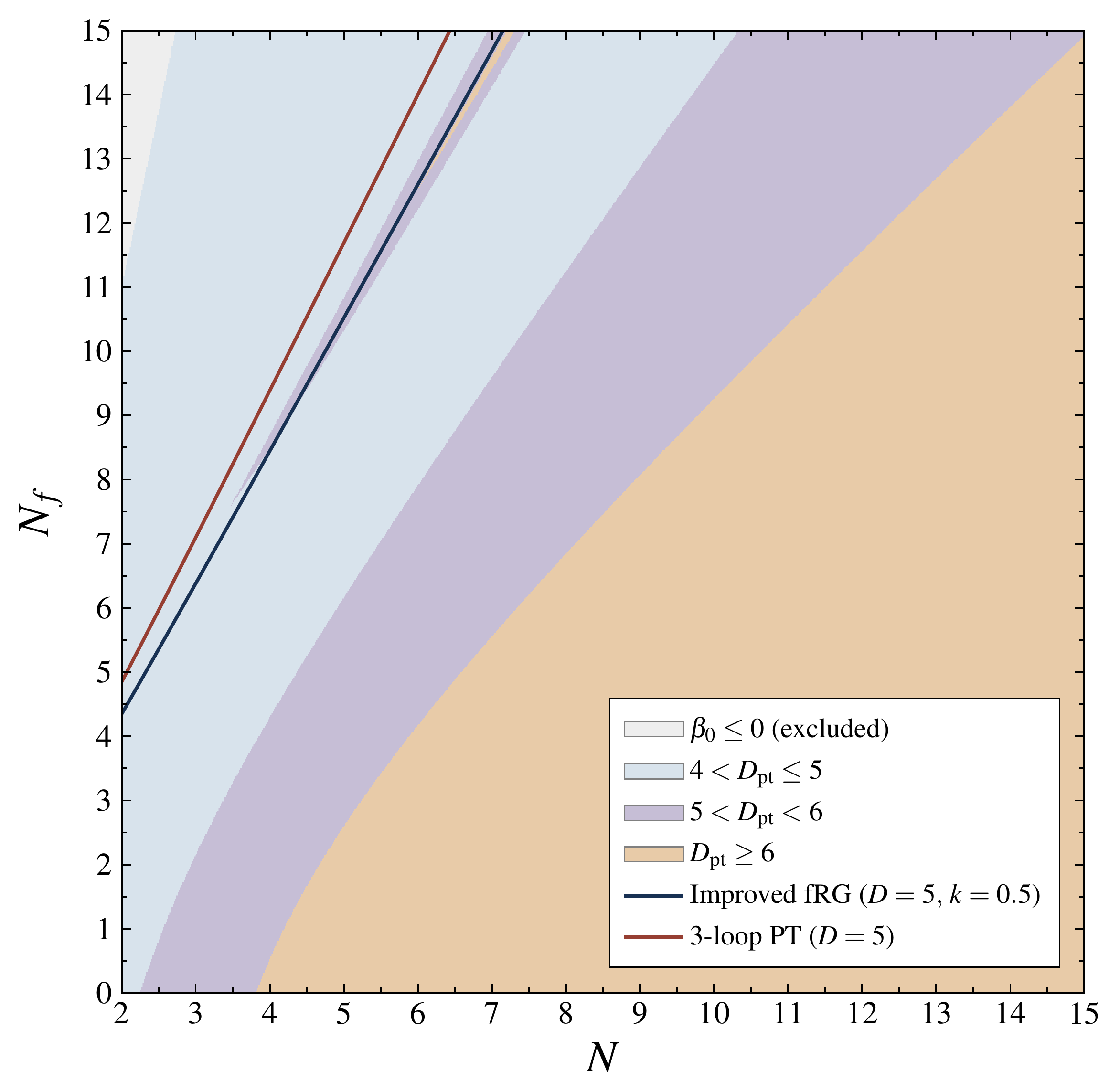}
    \caption{Perturbative-control dimension $\Dpt$ for fundamental $\SU(N)$ fermions,
using the one-coupling $F^2$ projection and $\theta=1/2$.
Light gray denotes $\beta_0\le0$, where the selected Gaussian-connected
positive UV branch is absent. Pale blue-gray, lavender, and sand denote
$4<\Dpt\le5$, $5<\Dpt<6$, and $\Dpt\ge6$, respectively.
Sand indicates that the hierarchy test has not failed below $D=6$; it does
not establish control of the full theory there.
Dark-blue and brick-red curves are candidate three-loop merger boundaries
from the rational ansatz ($k=1/2$) and the unresummed series, respectively.
The narrow band reflects suppression of an adjacent root increment;
passing this finite-order test need not persist at five loops.}
    \label{fig:phase_diagram}
\end{figure}

Applying this criterion across the $(N,N_f)$ plane yields the map shown in
Fig.~\ref{fig:phase_diagram} which covers only fermions in fundamental representation. The corresponding results for fermions in the adjoint and two-index representations are presented in Fig.~\ref{fig:beta_function_representations} of the Appendix. The map distinguishes four regions. In the light-gray region, $\beta_0=(11N-2N_f)/3\le0$, so the
one-loop beta function has no positive interacting fixed point
for $D>4$. These points are therefore excluded from the hierarchy
test, which follows the weak-coupling branch
$a_*\sim\epsilon/\beta_0$ from four dimensions.
This exclusion does not rule out fixed points generated by
higher-order terms.
In the sand-color region, at small $N_f/N$, the first crossing of the
baseline hierarchy threshold occurs at $D=6$ or higher, corresponding to $\Dpt\ge6$.  In the lavender-color region, $5<\Dpt<6$: the $D=5$ fixed point satisfies the baseline hierarchy criterion with $\theta=1/2$, while the first threshold crossing occurs between five and six dimensions. In the pale blue-gray region, $4<\Dpt\le5$, the threshold is reached at or below $D=5$. This classification concerns the loop hierarchy within the one-coupling
$F^2$ projection and does not assess the effects of omitted operators. For five-dimensional applications, the key boundary is the main $\Dpt=5$ contour separating the lavender and pale blue-gray regions, which delimits the low-flavor domain satisfying our baseline
criterion at $D=5$.

Two broad trends characterize the main regions of the map: $\Dpt$ generally decreases with $N_f$ at fixed $N$ and increases with $N$ at fixed $N_f/N$. The former trend is consistent with the decrease of $\beta_0=(11N-2N_f)/3$, which raises the leading-order estimate
$a_*\simeq\epsilon/\beta_0$ at fixed $\epsilon$ and tends to enhance higher-loop corrections. The latter trend reflects the sizable finite-$N$ enhancement of the
four-loop contribution in the main low-flavor region. Although finite-$N$ corrections start at order $1/N^2$ and already appear at two loops, they can be particularly pronounced at four loops: for $N_f/N=1.5$, $\beta_3/N^4\simeq78.1$ at $N=3$, compared with
$44.0$ in the planar limit. Where $R_4=|c_4/c_3|$ controls the hierarchy threshold, this enhancement tends to raise the ratio and bring its first threshold crossing to
smaller $\epsilon$, thereby lowering $\Dpt$ at small $N$. As $N$ increases, the enhancement weakens, allowing the hierarchy to persist to larger $\epsilon$ in these regions. Neither trend is strictly monotonic, since the hierarchy depends on
all retained loop coefficients. At larger $N$, scaling at fixed $N_f/N$ accounts for the approximately linear region boundaries. The main $\Dpt=5$ contour runs from $N_f\simeq2.2$ at $N=3$ to $N_f/N\simeq1.5$ at $N=15$.

Two consequences are worth stating explicitly. First, at $D=5$, no displayed $\SU(2)$ cell passes the baseline hierarchy criterion with $\theta=1/2$; even pure Yang--Mills theory gives $\Dpt\simeq4.87$. Failing this hierarchy test does not exclude a UV fixed point.
For five-dimensional pure $\SU(2)$ Yang--Mills theory, indications
of such a fixed point have been obtained from $\epsilon$-expansion
analyses~\cite{morrisRenormalizableExtradimensionalModels2005,DeCesare:2021}
and functional renormalization-group
calculations~\cite{giesRenormalizabilityGaugeTheories2003,pastor-gutierrezUVCompletionExtradimensional2023}.
Second, the estimated three-loop conformal-window boundaries (dark-blue and brick-red curves) run through the pale blue-gray region: the annihilation takes place where perturbative control at $D=5$ has already been lost, and only for $\theta\to1$
(Fig.~\ref{fig:threshold_sensitivity}) does the $\Dpt=5$ edge move up to the window
boundary.
The narrow band (See Table~\ref{tab:hierarchy_merger}) along the dark-blue merger curves reflects cancellations among higher-loop contributions that suppress $c_4$ relative to $c_3$, lowering $R_4=|c_4/c_3|$;
the other ratios also restrict its extent. In particular, $c_4$ vanishes exactly at a zero of $\beta_3$ on a common regular branch. The relevant lower-flavor zero moves from $N_f/N\simeq2.38$ at $N=4$ to $N_f/N\simeq2.04$ in the planar limit, passing through the merger region near $N_f/N\simeq2.09$. However, the small four-loop increment does not by itself establish stability at the next order. Extending the unresummed fixed-point calculation with the known five-loop coefficient~\cite{herzogFiveloopBetaFunction2017} illustrates this limitation. At $D=5$ in the planar limit, and at the flavor ratio selected by the three-loop rational merger, $N_f/N\simeq2.0913$, we find $R_4\simeq0.151$ but $|c_5/c_4|\simeq10.75$, where $c_5\equiv a_*^{(5)}-a_*^{(4)}$. Thus the small four-loop correction is not followed by a smaller five-loop correction in this example, even though the interacting UV fixed point persists.

The map is specific to the chosen $\MSbar$ continuation and hierarchy
threshold. The four-dimensional scheme independence of $\beta_0$ and $\beta_1$
does not make $\Dpt$ universal: for $D>4$, coupling redefinitions
also mix in the canonical term. For $\SU(N)$ with $N\geq3$ and odd $N_f$, the parity anomaly requires a half-odd-integer bare Chern--Simons level
\cite{Intriligator.KFive-Dimensional-Supersymmetric-Gauge1997};
the corresponding cells of Fig.~\ref{fig:phase_diagram} omit its
effects on the flow. Four-component Dirac spinors are compatible with $D=5$, but this
does not ensure equality between the continued $\MSbar$ coefficients
and those obtained from a direct five-dimensional calculation.

\section{Conformal window at $D=5$ from a fRG-informed resummation}
\label{sec:window}

We now turn to five dimensions and investigate whether fermion screening can eliminate an interacting UV fixed point through a merger with an interacting IR fixed point. 
The preceding hierarchy test does not establish perturbative control
throughout the candidate fixed-point region, motivating a construction
informed by the functional renormalization group (fRG).
The fRG flow of the
scale-dependent effective action $\Gamma_k$ is governed by the Wetterich equation
\cite{wetterichExactEvolutionEquation1993},
\begin{equation}
	\partial_{t}\Gamma_{k}=\frac{1}{2}\mathrm{STr}\left[\partial_{t}R_{k}\cdot
(\Gamma_{k}^{(2)}+R_{k})^{-1}\right],
\end{equation}
with $\partial_t=k\partial_k$, $\Gamma_k^{(2)}$ is the Hessian with
respect to the fluctuating fields, and the supertrace includes a minus
sign for Grassmann fields. The regulator $R_k$ suppresses low-momentum fluctuations. 
In the leading background-field approximation, the gauge sector is
\begin{equation}
 \Gamma_{\rm gauge}=\frac{Z_\kappa}{4}
 \int\!\dd^Dx\;F^a_{\mu\nu}\,F^{a\mu\nu},
 \label{eq:truncation_scheme}
\end{equation}
supplemented by covariant fermion kinetic, gauge-fixing, and ghost terms
\cite{pastor-gutierrezPhaseStructureExtradimensional2024,pastor-gutierrezUVCompletionExtradimensional2023}.
$Z_\kappa$ normalizes the background field strength. With $\bar g^2=k^{4-D}Z_k\,g^2$ and $\bar g$ held fixed,
the quantum contribution to the running of the dimensionless
coupling is encoded in the anomalous dimension
$\eta_g\equiv-\partial_t\ln Z_k$ and gives:
\begin{equation}
 \beta^{(D)}(a)=\frac12\partial_ta
 =\frac12(D-4+\eta_g)\, a.
 \label{eq:beta_frg}
\end{equation}
A nonzero fixed point in five dimensions satisfies $\eta_g=-1$. In this truncation, we identify the background field coupling with the fermion-gluon and three-gluon vertex couplings.
We do not resolve their separate flows or the full constraints from modified Slavnov-Taylor and background field identities \cite{ellwangerHirschWeber1996,freireLitimPawlowski2000}. 

For massless fundamental Dirac fermions, neglecting fermion and ghost anomalous dimensions gives, in the normalization of Ref.~\cite{pastor-gutierrezPhaseStructureExtradimensional2024},
\begin{equation}
 \eta_g^{(0)}=-\frac{Aa}{1-Ca},\qquad
 A=7N-\frac43N_f,\quad C=\frac{19}{6}N.
 \label{eq:first_order_eta}
\end{equation}
The adjoint flavor extension replaces $A$ by $A-8N \, N_{\rm ad}/3$.
The main analysis sets $N_{\rm ad}=0$. The numerator separates gauge anti-screening from fermion screening. The denominator follows from the implicit equation:
\begin{equation}
 \eta_g^{(0)}=-Aa+Ca\,\eta_g^{(0)}.
 \label{eq:feedback}
\end{equation}
It resums the feedback caused by the running normalization inside $\partial_tR_\kappa$, rather than all gauge interactions \cite{giesRunningCouplingYangMills2002,giesIntroductionFunctionalRG2012}.
For $A>0$, it drives the beta function downward toward a pole at $a_{\rm pole}=1/C$. On the regular branch below that pole, the only interacting zero is $a_*=1/(A+C)$; it cannot annihilate with an IR partner. Increasing $N_f$ moves this root towards the pole as $A\to0^+$. The pole is an artifact of the approximation, not a physically motivated singularity. The feedback derivation, threshold normalization, and the dimensionless measure $\xir=Ca$ are discussed in Appendix~\ref{app:feedback}.

Higher gauge operators can change this strong-coupling continuation. In a Gies-type treatment, a function $W_\kappa(\vartheta)=Z_\kappa\vartheta+\sum_{n\ge2}w_{n,\kappa}\vartheta^n$ with $\vartheta=F^a_{\mu\nu}F^{a\mu\nu}/4$ incorporates an infinite tower
of field-strength invariants with running coefficients $w_{n,\kappa}$. These additional flows can soften the pole-driven decrease and allow an IR partner \cite{giesRenormalizabilityGaugeTheories2003}. After aligning weak-coupling normalizations, one may therefore expect a more complete flow to interpolate between continuations dominated by pole feedback and by higher-operator effects over the relevant coupling range.
This motivates retaining the downward anomalous-dimension feedback while allowing higher-order terms to turn the beta function upward. To explicitly incorporate known perturbative information, we take the rational structure of the leading $F^2$ flow as a tractable starting point, regularize its pole, and determine the numerator by matching to the $\MSbar$ series at successive orders. Keeping the denominator fixed then allows to test how the candidate fermionic merger boundary responds to each additional matching order. The higher-operator results motivate this construction, but do not determine its coefficients or provide numerical bounds on the merger curve.

We implement this idea by retaining a rational anomalous dimension while
matching its weak-coupling expansion to known $\MSbar$ coefficients:
\begin{align}
 \eta_g^{\rm imp}(a)&=-\frac{2\beta_0a+b_1a^2+b_2a^3}{Q_k(a)},
 \label{eq:improved_eta}\\
 Q_k(a)&=1-Ca+kC^2a^2,\qquad k>\frac14,
 \label{eq:denominator}\\
 b_1&=2(\beta_1-C\beta_0),\nonumber\\
 b_2&=2(\beta_2-C\beta_1+kC^2\beta_0).
 \label{eq:matching}
\end{align}
The linear term $-Ca$ retains the leading regulator-feedback structure;
the quadratic term removes the positive-coupling pole. The numerator is
then fixed by matching, so the construction preserves every known
coefficient through three loops without double counting that feedback.
It is a $\MSbar$-anchored, fRG-informed completion, not a rigorous solution of the higher-operator flow. We refer to it as the improved fRG ansatz in
this restricted sense.

Our baseline is $k=1/2$, for which
$Q_{1/2}=[(Ca-1)^2+1]/2>0$: the minimum lies at the original pole.
This simple geometric prescription does not introduce fitted merger input.
Positivity removes the spurious singularity but does not enforce an IR
root; that root, if present, follows from the matched numerator.
Appendix~\ref{app:matching} gives the matching derivation and variations in $k$.

\begin{figure*}[t]
    \centering
    \includegraphics[width=0.48\textwidth]{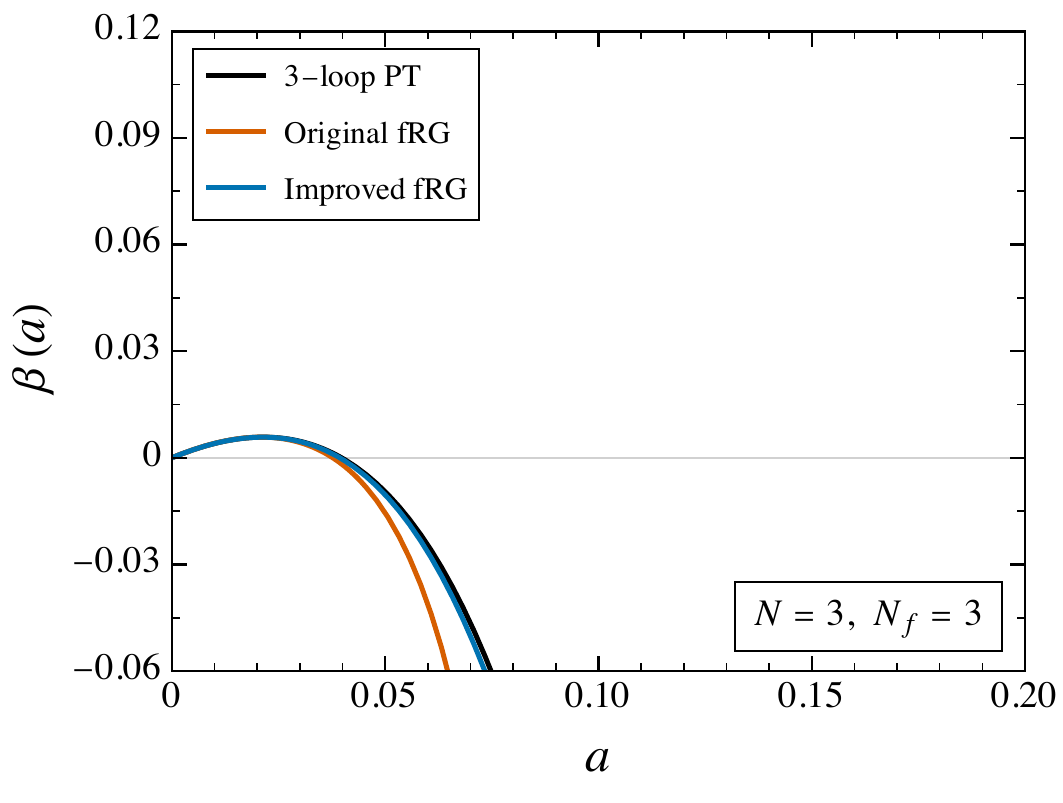}
    \hfill
    \includegraphics[width=0.48\textwidth]{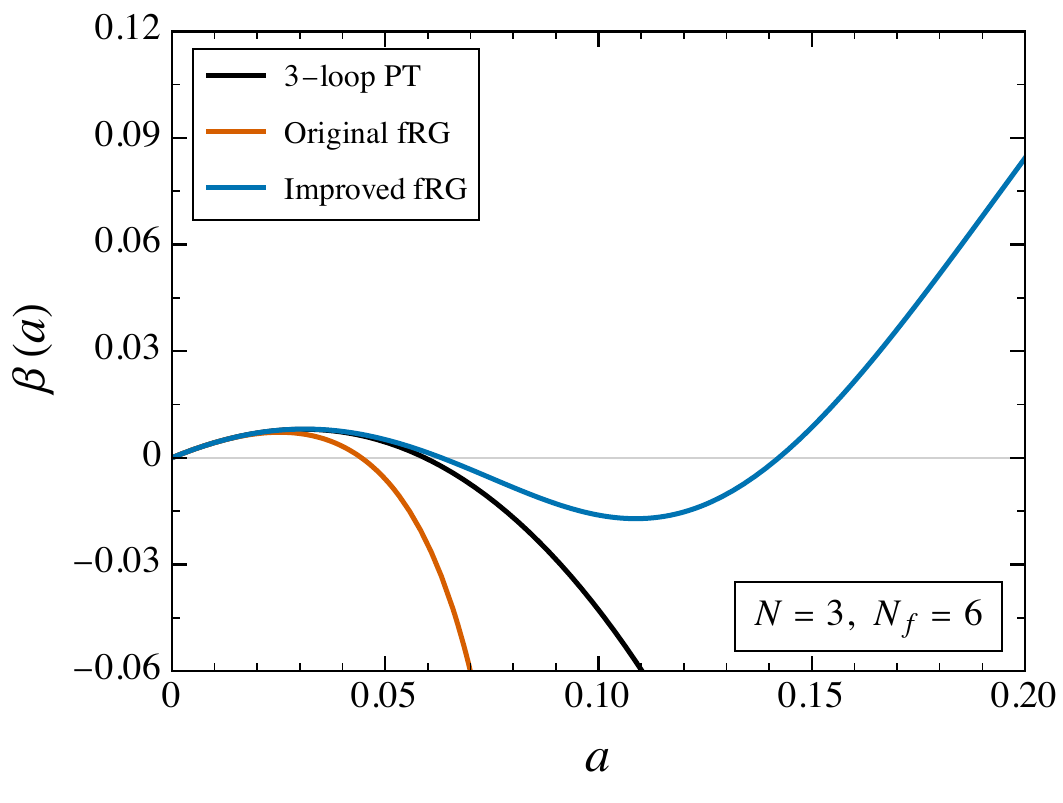}
    \vspace{1em}
    \includegraphics[width=0.48\textwidth]{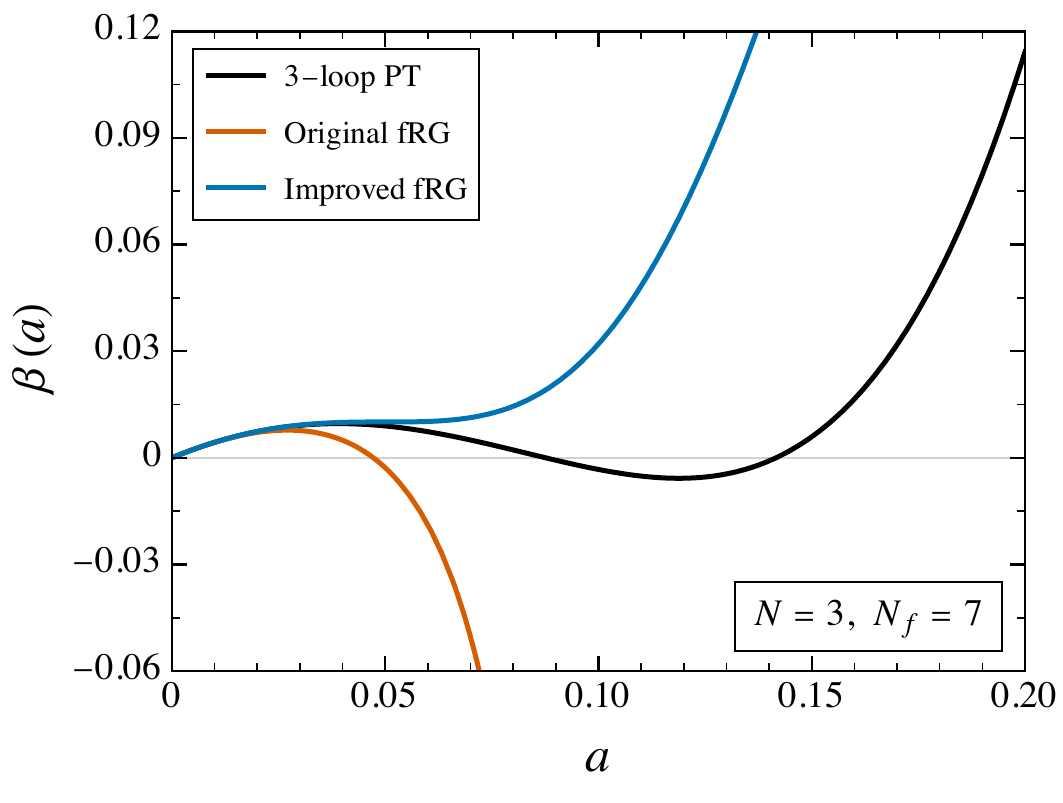}
    \hfill
    \includegraphics[width=0.48\textwidth]{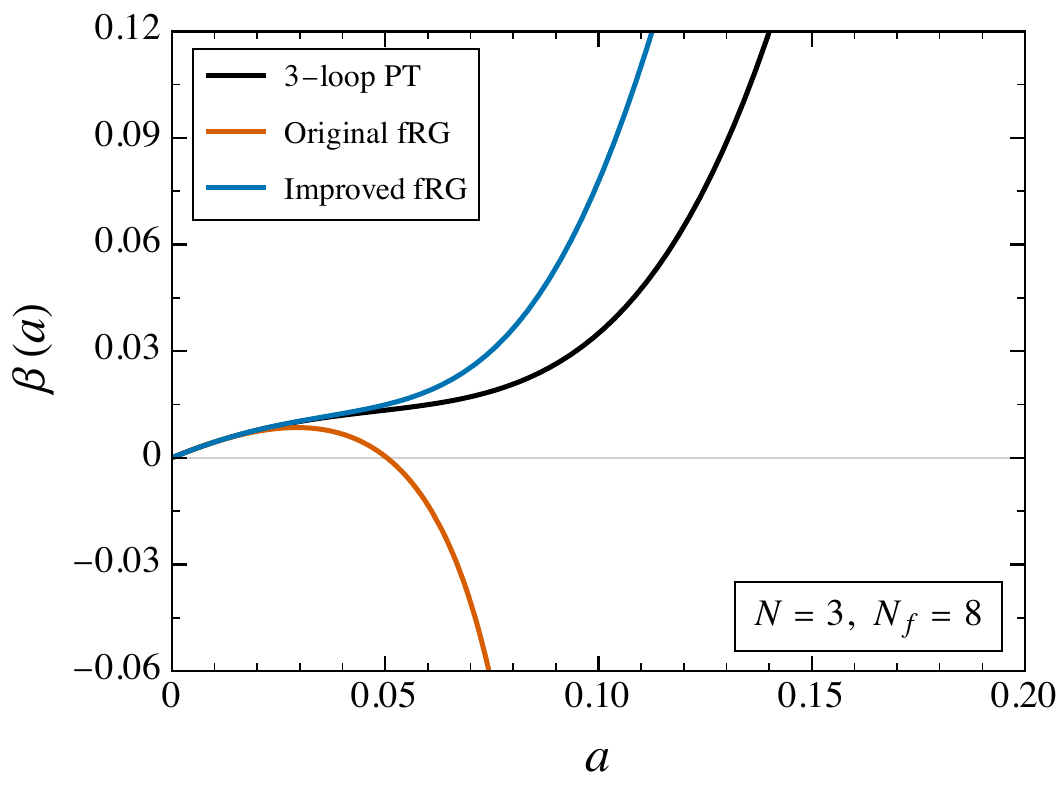}
    \caption{Gauge-coupling beta function at $D=5$ for $\SU(3)$ with fundamental fermions
    at representative $N_f$: three-loop perturbation theory, the leading fRG result
    \eqref{eq:first_order_eta}, and the improved beta function obtained from
    Eq.~\eqref{eq:improved_eta} with $k=1/2$. The UV fixed point and its IR partner
    (the first and second positive zeros) merge between $N_f=6$ and $7$ in the improved beta function
    and between $N_f=7$ and $8$ at three loops. We formally identify the threshold-normalized coupling with the $\MSbar$ coupling in this comparison. No explicit scheme conversion has been performed.}
    \label{fig:beta_function_plot}
\end{figure*}

For $\SU(3)$, the illustrative curves in Fig.~\ref{fig:beta_function_plot}
show a UV--IR merger between $N_f=6$ and $7$ in the three-loop matched
ansatz, compared with between $7$ and $8$ in the unresummed three-loop
series. The merger satisfies
\begin{equation}
 \beta^{(5)}(a_*,\Nfcr)=0,\qquad
 \partial_a\beta^{(5)}(a_*,\Nfcr)=0.
 \label{eq:merger}
\end{equation}
We treat $N_f$ continuously to locate this boundary; physical flavor
multiplicities remain integers. With $N_f\propto N$ and $a\propto1/N$,
the baseline three-loop rational estimate at $k=1/2$ approaches
$\Nfcr/N\simeq2.09$ and $Na_*\simeq0.26$ at large $N$.
It is approximately a $10\%$ displacement from the unresummed boundary,
$\Nfcr/N\simeq2.32$, and measures the sensitivity to the assumed higher-order
completion. Within this three-loop rational family, varying $k$ from
$(1/4)^+$ to $1/2$ changes $\Nfcr$ by approximately $1\%$ and the merger
coupling $a_*$ by approximately $8\%$, relative to the $k=1/2$ baseline.
Tables~\ref{tab:main} and \ref{tab:critical_Nf} give the finite-$N$
results in Appendix~\ref{app:tables}.

The same denominator makes a matching-order test possible without changing
the resummation prescription:
\begin{align}
 \eta_g^{[L]}(a)&=-\frac{2a}{Q_{1/2}(a)}
                  \sum_{j=0}^{L-1}h_j a^j,\quad L=1,\ldots,4,
 \nonumber\\
 h_j&=\beta_j-C\beta_{j-1}+\frac{C^2}{2}\beta_{j-2},
 \label{eq:matching_order_family}
\end{align}
where $\beta_{-1}=\beta_{-2}=0$ and
$\beta_{[L]}^{(5)}=a(1+\eta_g^{[L]})/2$.
The bracket $[L]$ counts matched orders, distinguishing these rational
flows from the unresummed $L$-loop roots $a_*^{(L)}$.
The $L=1$ member already contains a quadratic denominator and is not
Eq.~\eqref{eq:first_order_eta}.
As shown in Table~\ref{tab:matching_order}, the $\SU(3)$ merger
shifts by only $0.696\%$ in $\Nfcr$ and $0.726\%$ in $a_*$
between $L=2$ and $3$, compared with the respective
one-to-two-loop shifts of $52.585\%$ and $41.420\%$ respectively.
This marked reduction indicates low-order stabilization within the chosen rational family, lending support to the merger as a candidate upper flavor boundary for five-dimensional asymptotic safety.
The analytic lower-order solutions and the color dependence are given in the Appendix.

\begin{table}[t]
\caption{Matching-order dependence of the $D=5$ merger for $\SU(3)$ with
fundamental Dirac fermions and the same $Q(a)$ at $k=1/2$.
The last two columns give the successive relative changes
$\delta_X^{[L]}=100|X^{[L]}-X^{[L-1]}|/|X^{[L-1]}|$;
these are order shifts, not statistical uncertainties.}
\label{tab:matching_order}
\setlength{\tabcolsep}{4.3pt}
\begin{ruledtabular}
\begin{tabular}{ccccc}
$L$ & $N_f^{\mathrm{cr}[L]}$ & $a_*^{[L]}$ & $\delta_{N_f}^{[L]}\;(\%)$ & $\delta_a^{[L]}\;(\%)$\\
\hline
1 & 13.548728 & 0.148865 & --- & ---\\
2 &  6.424145 & 0.087205 & 52.585 & 41.420\\
3 &  6.379443 & 0.086572 &  0.696 &  0.726\\
\end{tabular}
\end{ruledtabular}
\end{table}

We next examine whether this candidate boundary lies within the
perturbative-control region identified from the unresummed loop
expansion. Figure~\ref{fig:phase_diagram} overlays the merger curve
on the loop-hierarchy map. At the $\SU(3)$ three-loop rational merger
flavor, $N_f\simeq6.379$, the unresummed fixed-point sequence at $D=5$
gives $(R_2,R_3,R_4)\simeq(0.16,0.27,1.01)$.
The four-loop increment therefore violates the baseline hierarchy
criterion, $R_4>\theta=1/2$.
The candidate boundary lies beyond the main low-flavor
perturbative-control region, although for some colors it intersects
the narrow passing band discussed above.

Four-loop matching probes whether the interacting UV branch can
persist beyond the three-loop merger boundary. Incorporating the
known four-loop coefficient:
\cite{herzogFiveloopBetaFunction2017} while retaining the same
denominator $Q(a)$ gives
\begin{align}
h_3&=\beta_3-C\beta_2+\frac{C^2}{2}\beta_1,\nonumber\\
\beta_{[4]}^{(5)}-\beta_{[3]}^{(5)}
&=-\frac{h_3a^5}{Q(a)}.
\label{eq:four_loop_matching_shift}
\end{align}
For $2\le N\le15$, numerical root analysis finds a single positive
UV zero throughout $0\le N_f/N\le11/2$, with no UV--IR merger.
The interacting UV branch therefore survives four-loop matching and extends beyond the three-loop boundary
(Fig.~\ref{fig:four_loop_uv_persistence} for $\SU(3)$), opening
a larger candidate domain of asymptotic safety within this
rational family. 
With the same denominator, five-loop matching yields both a UV
and an IR zero for $\SU(3)$ at the three-loop merger flavor
$N_f\simeq6.379$ (Appendix~\ref{app:matching}).
The IR zero thus reappears at five loops; determining whether
the two zeros merge requires tracking their evolution as $N_f$ varies.
The low-order stabilization of the merger coordinates supports
annihilation as a candidate mechanism for an upper flavor boundary,
whereas the four-loop continuation exhibits a UV branch extending
to higher flavor numbers without this termination.
The improved construction thus allows us to assess both a
merger-limited ultraviolet domain and its possible extension
under higher-order matching.

\section{Conclusions}

We have developed complementary perturbative and fRG-informed
diagnostics for exploring gauge-theory asymptotic safety above four
dimensions. Using the four-loop $\MSbar$ beta function continued to
$D=4+2\epsilon$, we map $\Dpt$ across color and flavor numbers,
identifying the domain that satisfies a specified hierarchy of
successive fixed-point shifts. At $D=5$, our improved beta function
combines a rational structure motivated by anomalous-dimension
feedback with pole regularization and matching to the known
perturbative coefficients. The resulting phase map relates the
main region satisfying the baseline hierarchy criterion to
candidate conformal-window boundaries, providing a quantitative
guide within these approximations for selecting matter content
in extra-dimensional models.

The one-to-three-loop matching sequence exhibits low-order
stabilization of the merger coordinates, yielding
$\Nfcr/N\simeq2.09$ at large $N$. This supports UV--IR annihilation
as a candidate mechanism limiting the flavor range of
five-dimensional asymptotic safety. 
Four-loop matching reveals a complementary possibility:
with the same denominator, the interacting UV branch persists
beyond the three-loop merger boundary, with no UV--IR merger
found over the color and flavor ranges studied.
Within the chosen rational family, this continuation extends
the candidate domain of asymptotic safety to larger flavor numbers.
At five-loop matching, an IR partner reappears alongside the UV
zero at the $\SU(3)$ three-loop merger flavor, $N_f\simeq6.379$.
The matching-order comparison
therefore exposes the sensitivity of the merger scenario while
identifying how higher-order information can alter the extent
of the UV branch. Variations of the completion and regulator
inputs further characterize the sensitivity of these estimates. The improved construction thus provides a common framework for examining both the persistence of the UV branch and its possible termination through fixed-point annihilation.

These results provide a concrete starting point for extending
the analysis to coupled gauge and matter flows. Higher gauge
operators and gauge-induced fermionic interactions can modify
the fixed-point structure and its stability
\cite{giesChiralPhaseStructure2006,
goertzGaugeFermionCartographyConfinement2024}.
Including their flows, together with a consistent treatment of
background and vertex couplings subject to the regulator-modified
identities, would test whether the candidate gauge fixed points
extend to fixed points of the coupled system with a
finite-dimensional UV critical surface. Our phase map identifies
the matter contents and parameter regions where these further
tests are most informative, connecting perturbative information
and fRG-motivated continuations to the search for viable
five-dimensional ultraviolet completions.

\begin{acknowledgments}
Acknowledgments— Zhi-Wei Wang, Jie Liu, Yu-Bo Zhao are partially supported by the National Natural Science Foundation of China (Grant No.~12475105, Grant No.~12622508). We thank Masatoshi Yamada for helpful discussions.
\end{acknowledgments}

\bibliography{refs}

\twocolumngrid

\appendix
\setcounter{secnumdepth}{1}
\section*{Appendix}
\section{Loop hierarchy, branch selection, and narrow passing bands}
\label{app:hierarchy}
\begin{figure*}[tp]
\centering
\includegraphics[width=0.90\textwidth]{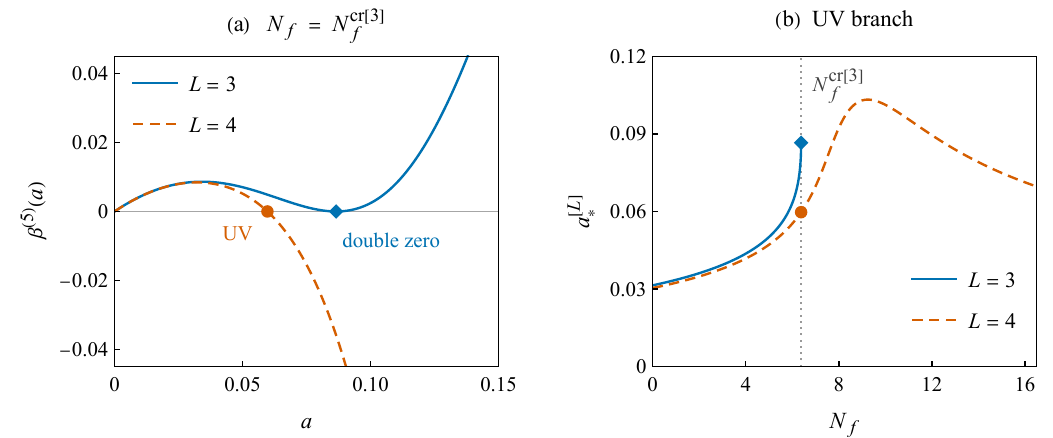}
\caption{Effect of four-loop matching at $D=5$ for $\SU(3)$ with the same
regularized denominator, $k=1/2$. (a) At the three-loop merger flavor
$N_f=6.379443$, the three-loop double zero is replaced by a simple
four-loop UV zero. (b) The four-loop UV branch continues through the
three-loop boundary. Blue solid and orange dashed curves denote three-
and four-loop matching, respectively. These are zeros of the specified
one-coupling ansatz, not a test of the complete fixed-point system.}
\label{fig:four_loop_uv_persistence}
\end{figure*}

For fixed $(N,N_f)$, write
$P_L(a)=\sum_{n=0}^{L-1}\beta_na^{n+1}$, so that the unresummed root
equation is $\epsilon=P_L(a)$.
At an interacting zero,
\begin{equation}
 \partial_a\beta^{(4+2\epsilon)}(a_*)=-a_*P_L'(a_*),
 \label{eq:root_slope}
\end{equation}
and the branch with $P_L'>0$ is UV attractive in the gauge direction.
For $N=3$, $N_f=7$,
$P_3=(19/3)a+(40/3)a^2-(12629/54)a^3$.
Solving $P_3'=0$ gives $a\simeq0.11590$ and
$P_3(a)\simeq0.54903$; the two roots merge at this maximum.
The positive four-loop coefficient, $\beta_3\simeq982.56$, makes $P_4$
monotonically increasing for $a>0$ for this benchmark, and restores a
unique positive root for every $\epsilon>0$ within that truncation.
This is a statement about the truncated polynomial, not about arbitrary
dimensions in the complete theory.

\begin{figure*}[tp]
\centering

\EMpanel{figures/phase_diagram_0.5.png}
        {(a) $\theta=0.5$}{0.25}
\hfill
\EMpanel{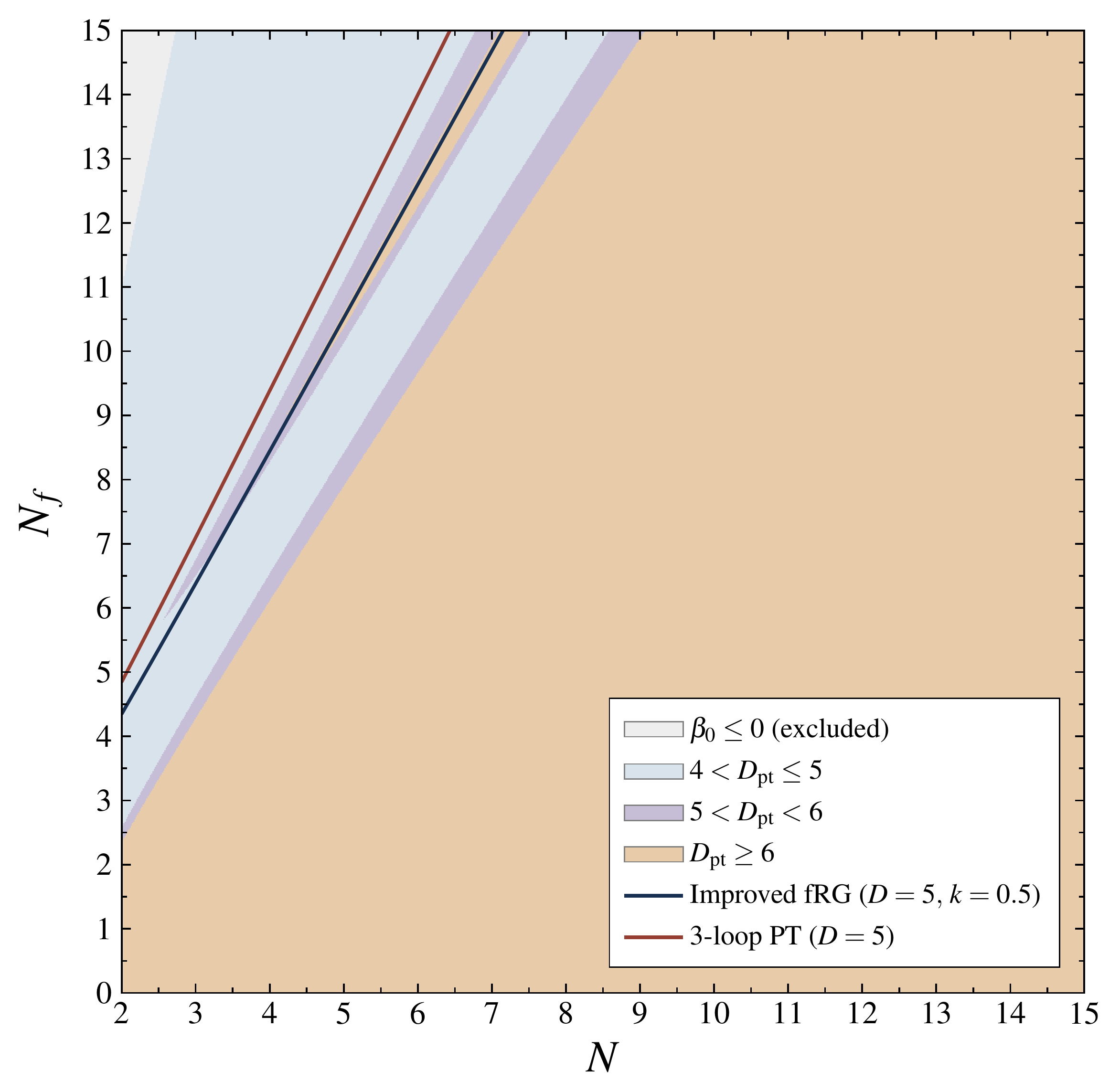}
        {(b) $\theta=0.8$}{0.25}
\hfill
\EMpanel{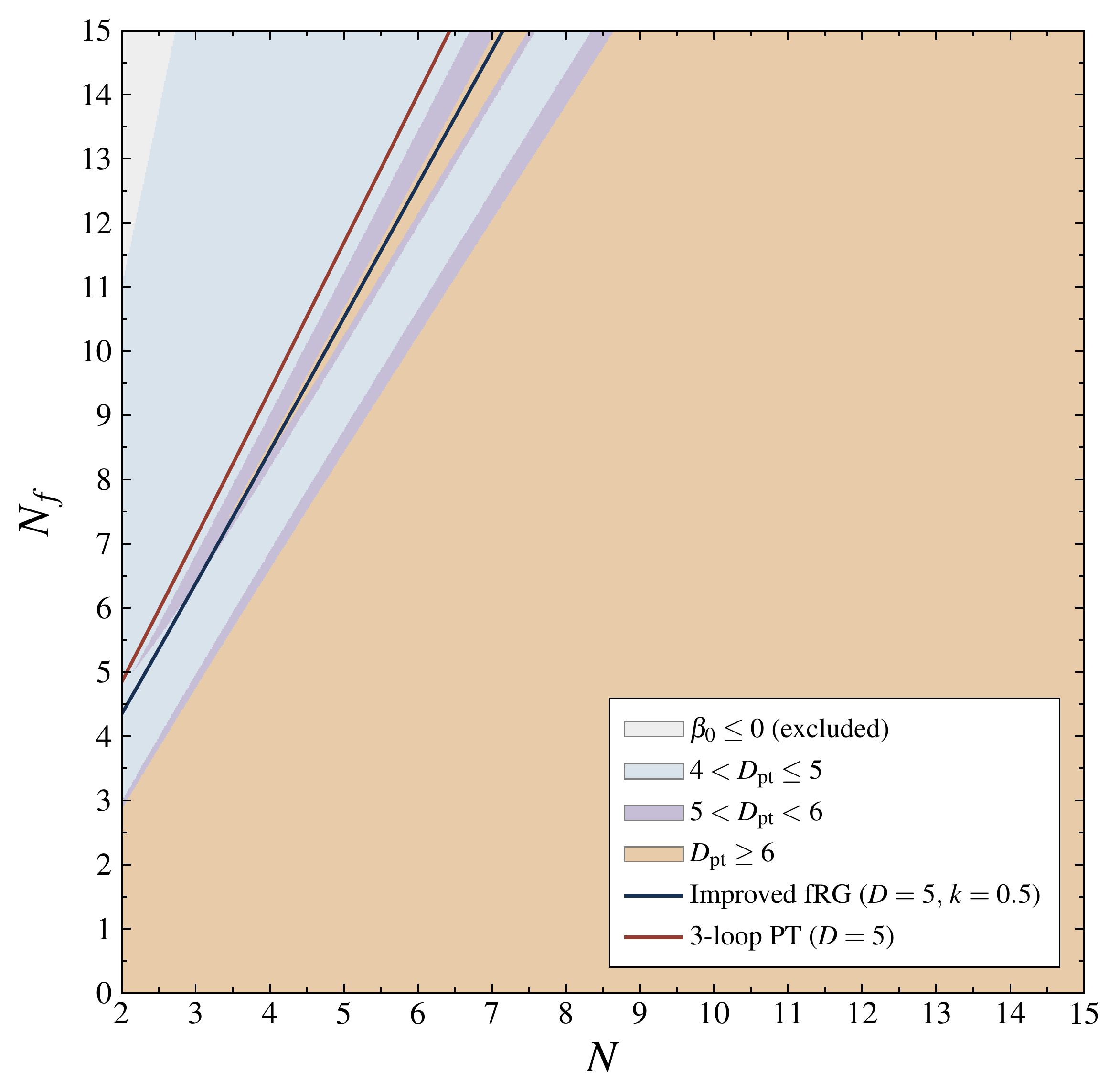}
        {(c) $\theta=1$}{0.25}
\par

\caption{%
Perturbative-control dimension $\Dpt$ in the $(N,N_f)$ plane
for (a) $\theta=0.5$, (b) $\theta=0.8$, and (c) $\theta=1$.
The unfiltered color coding and treatment of narrow passing regions follow Fig.~\ref{fig:phase_diagram}. White indicates termination of a weak-coupling branch before a ratio crossing, or an undefined ratio.
Dark-blue and brick-red curves show candidate $D=5$ merger boundaries from the three-loop rational ansatz with $k=1/2$ and the unresummed three-loop beta function, respectively.}

\label{fig:threshold_sensitivity}
\end{figure*}

For this example, the first crossing is $R_3=1/2$ at
$\epsilon\simeq0.20$; $R_4$ reaches the threshold near
$\epsilon\simeq0.49$. At $N_f=2$, instead,
$R_4(1/2)\simeq0.495$ and its first crossing occurs near
$\epsilon\simeq0.516$, giving $\Dpt\simeq5.03$.
Figure~\ref{fig:threshold_sensitivity} displays the change of the map
when $\theta$ is relaxed from $1/2$ to $0.8$ and $1$.

At small $\epsilon$, the adjacent increments obey
\begin{equation}
 c_L=-\frac{\beta_{L-1}}{\beta_0^{L+1}}\epsilon^L
      +O(\epsilon^{L+1}),\qquad L\ge2.
 \label{eq:increment_small_epsilon}
\end{equation}
An exact relation makes the cancellation mechanism transparent.
Because $P_L(a_L)=P_{L-1}(a_{L-1})=\epsilon$, where
$a_L\equiv a_*^{(L)}$, the mean-value theorem gives
\begin{equation}
 c_L=-\frac{\beta_{L-1}a_L^L}{P_{L-1}'(\zeta_L)},
 \label{eq:increment_identity}
\end{equation}
for a point $\zeta_L$ between the two roots, provided the indicated
derivative does not vanish. Equivalently one can use a polynomial divided
difference, avoiding subtraction of nearly equal roots.
Thus $\beta_3=0$ makes $c_4=0$ on a common regular branch, and a nearby
suppression can make $R_4$ small without suppressing the next order.
The lower-flavor zero of $\beta_3$ moves from $N_f/N\simeq2.38$ at $N=4$
to $\simeq2.04$ in the planar limit, close to the rational merger region.
The other ratios still determine the extent of the passing band; it need
not be centered on the coefficient zero.

The opposite problem occurs when the denominator of a ratio vanishes.
For $\SU(3)$,
$\beta_2=2857/2-5033N_f/18+325N_f^2/54$ is zero at
$N_f\simeq5.843996$. Then $c_3=0$, so $R_4$ is undefined although
$c_4/a_*^{(1)}=O(\epsilon^3)$ tends to zero.
Undefined ratios are masked rather than interpreted as a physical loss
of the fixed point. Likewise, termination before any crossing is distinct
from the ratio-defined $\Dpt$.

Table~\ref{tab:hierarchy_merger} evaluates the unresummed roots at the
flavor numbers selected by the three-loop rational merger; the rational
merger coupling itself is not used in these ratios. Several colors pass
the four-loop test there. A known next-order check shows why this is not
enough: at $D=5$ in the planar limit with $N_f/N\simeq2.0913$, adding the
five-loop coefficient \cite{herzogFiveloopBetaFunction2017} gives
$R_4\simeq0.151$ but $|c_5/c_4|\simeq10.75$, with
$c_5=a_*^{(5)}-a_*^{(4)}$. The positive UV root persists. This example
diagnoses a failure of the successive-increment hierarchy, not the
disappearance of the fixed point or a proof of divergence of the full series.

The finite-$N$ mechanism discussed in the main text is conditional.
Holding $\beta_0,\beta_1,\beta_2$ and $\epsilon$ fixed, implicit differentiation
on a regular four-loop UV branch yields
\begin{equation}
 \frac{\partial a_*^{(4)}}{\partial\beta_3}
 =-\frac{(a_*^{(4)})^4}{P_4'(a_*^{(4)})}<0.
 \label{eq:fourloop_response}
\end{equation}
If $\beta_3>0$ and both roots lie on the increasing continuation of $P_3$,
then $a_*^{(4)}<a_*^{(3)}$, so increasing $\beta_3$ increases $|c_4|$
at fixed $c_3$. This raises $R_4$ and can advance its threshold crossing.
Actual variation of $N$ changes the lower coefficients too, so this sign
argument explains a tendency in the indicated region rather than a
general ordering of $\Dpt$.

\section{Regulator feedback and the original pole scale}
\label{app:feedback}
The origin of Eq.~\eqref{eq:feedback} can be seen from a regulator of the
form $R_\kappa(p^2)=Z_\kappa p^2r(y)$, with $y=p^2/\kappa^2$ and
dimensionless shape function $r$. At fixed momentum,
$\partial_ty=-2y$ and $\partial_tZ_\kappa=-\eta_g Z_\kappa$ give
\begin{equation}
 \partial_tR_\kappa
 =Z_\kappa p^2[-\eta_g r(y)-2yr'(y)].
 \label{eq:regulator_derivative}
\end{equation}
The first term feeds the anomalous dimension back into its own projected
flow. Solving the resulting linear equation resums powers of $Ca$.
In a calculation that distinguishes background and fluctuation fields,
the normalization in a fluctuation regulator and the corresponding
anomalous dimension need not equal the background $Z_\kappa$ and
$\eta_g$. Their identification here belongs to the leading closure.
A regulator adjusted to the full Hessian can additionally depend on
running higher-operator couplings; this is more information than the
single feedback in Eq.~\eqref{eq:feedback}
\cite{giesIntroductionFunctionalRG2012}.

The quoted leading flow uses the threshold-normalized coupling of
Ref.~\cite{pastor-gutierrezPhaseStructureExtradimensional2024}.
Its gluonic coefficient $7N$ differs from $22N/3$ in the four-dimensional
$\MSbar$ continuation. The massless fundamental-fermion coefficient
$4N_f/3$ agrees for these two prescriptions. Neither agreement nor
disagreement alone defines a full scheme transformation in five
dimensions: the dimension-dependent field trace, threshold integrals,
and coupling normalization all matter. Accordingly, importing the linear
denominator coefficient into an $\MSbar$-matched numerator is an ansatz.

We retain the original pole scale as a useful coordinate,
\begin{equation}
 \xir\equiv Ca=\frac{a}{a_{\rm pole}},\qquad
 a_{\rm pole}=\frac{6}{19N}.
 \label{eq:xi}
\end{equation}
For $0\le\xir<1$, the leading unregularized expression has the geometric
enhancement $1/(1-\xir)$: $\xir=0.1$ gives an $11\%$ increase and
$\xir=1/2$ a factor of two. These factors quantify that denominator alone.
They are not relative errors on the improved beta function, the fixed-point
coupling, or a physical observable.

Along the main $\Dpt=5$ contour, evaluating $\xir$ at the unresummed
four-loop root gives about $0.34$ for $N=3$ and $0.43$ for $N=15$.
The three-loop rational merger instead has $\xir\simeq0.82$, for which
$Q_{1/2}\simeq0.52$. The compared couplings come from different
approximations. Table~\ref{tab:xi} gives dimensions at which the four-loop
root reaches smaller specified values of $\xir$; its entries constrain
only this feedback measure, not the total correction to the root.

\section{Matching formulas, merger conditions, and sensitivity tests}
\label{app:matching}
For a general denominator $Q_k=1-Ca+kC^2a^2$, matching through $L$ loops
means imposing
$-\eta_g^{[L]}=2\sum_{j=0}^{L-1}\beta_ja^{j+1}+O(a^{L+1})$.
Multiplication by $Q_k$ then gives
\begin{equation}
 h_j(k)=\beta_j-C\beta_{j-1}+kC^2\beta_{j-2},
 \label{eq:matching_general_k}
\end{equation}
which determines the numerator uniquely at fixed $Q_k$ and $L$.
The feedback terms in the expansion of $1/Q_k$ are compensated in these
matched coefficients; they are not added again to the perturbative data.
The ansatz predicts a particular tail beyond the matched order, rather
than deriving that tail from the exact fRG flow. In particular, an IR
partner in the $L=1$ rational member is generated by its denominator
completion; the unresummed one-loop polynomial has no such partner.

For $k>1/4$, $Q_k$ is positive for every real $a$. At the endpoint,
$Q_{1/4}=(1-Ca/2)^2$ has a positive zero, so the endpoint itself is not
globally pole free. The notation $k_+$ in the tables means the limit
$k\to(1/4)^+$, for which the listed merger coordinates remain finite.
Our baseline $k=1/2$ is the largest $k$ for which $Q_k$ is nonincreasing
throughout $0\le Ca\le1$; its minimum coincides with the original pole.
This geometric motivation is not a uniqueness criterion for the completion.

\begin{figure*}[tp]
\centering
\includegraphics[width=0.90\textwidth]{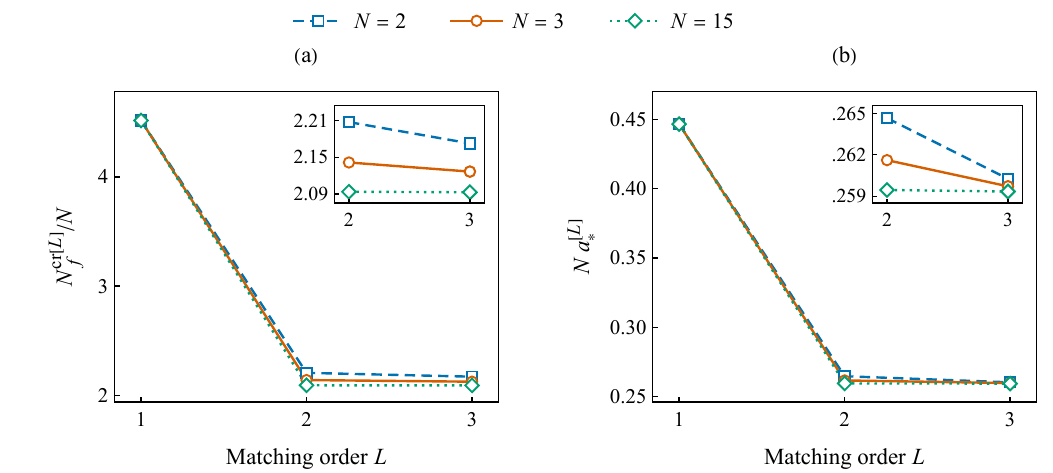}
\caption{Low-order stabilization of the rational merger at $D=5$ and $k=1/2$:
(a) critical flavor ratio $N_{f}^{\mathrm{cr}[L]}/N$ and (b) merger coupling $Na_*^{[L]}$
for $N=2,3,15$. All three matching orders use the same quadratic denominator.
The insets resolve the much smaller changes between $L=2$ and $3$.
Lines connect discrete matching orders and do not represent an extrapolation.
}
\label{fig:matching_order_stability}
\end{figure*}

\subsubsection*{Dependence on the matching order}

For fundamental Dirac fermions, introduce $x=N_f/N$ and $\lambda=Na$.
With $q(\lambda)=1-19\lambda/6+361\lambda^2/72>0$,
Eq.~\eqref{eq:matching_order_family} gives
$\beta_{[L]}^{(5)}=aF_L(\lambda,x)/q(\lambda)$, where
\begin{align}
F_{1,2}&=\tfrac12-B\lambda+A_{1,2}\lambda^2,\nonumber\\
F_3&=\tfrac12-B\lambda+A_2\lambda^2-\bar h_2\lambda^3,
\label{eq:matching_order_polynomials}\\
B&=\frac{21}{4}-\frac{2x}{3},\qquad A_1=\frac{361}{144},\nonumber\\
A_2&=\frac{401}{144}+\left(\frac{20}{9}-\frac{1}{N^2}\right)x,
\qquad \bar h_2=\frac{h_2}{N^3}.
\label{eq:matching_order_AB}
\end{align}
Thus the first two orders have a positive double root when $B^2=2A_L$ and
$B>0$. Writing $x^{\mathrm{cr}[L]}=N_f^{\mathrm{cr}[L]}/N$ and
$\lambda_*^{[L]}=Na_*^{[L]}$, their solutions are
\begin{align}
x^{\mathrm{cr}[1]}&=\frac{63-19\sqrt{2}}{8},
&\lambda_*^{[1]}&=\frac{6\sqrt{2}}{19},\nonumber\\
x^{\mathrm{cr}[2]}&=\frac{t_N-\sqrt{t_N^2-3167}}{8},
&\lambda_*^{[2]}&=\frac{1}{B(x^{\mathrm{cr}[2]})},
\label{eq:matching_order_closed}
\end{align}
with $t_N=103-18/N^2$.
For $L=3$ we solve $F_3=\partial_\lambda F_3=0$ numerically, tracking the
coalescence of the two smallest positive roots.

These solutions describe annihilation as $N_f$ increases.
For each of $L=1,2,3$ and $2\le N\le15$, the derivatives at the merger obey
$F_x>0$ and $F_{\lambda\lambda}>0$. Locally,
\begin{equation}
\lambda_\pm=\lambda_*\pm
\sqrt{\frac{2F_x}{F_{\lambda\lambda}}(x^\mathrm{cr}-x)}
+O(x^\mathrm{cr}-x),
\label{eq:matching_order_localfold}
\end{equation}
where derivatives are evaluated at $(\lambda_*,x^\mathrm{cr})$ for the chosen order.
Since $q>0$, the smaller root has $\partial_a\beta^{(5)}<0$ (UV) and the
larger has $\partial_a\beta^{(5)}>0$ (IR); the pair disappears for $x>x^\mathrm{cr}$.
This checks the merger of the gauge-coupling roots within each rational model.

Fig.~\ref{fig:matching_order_stability} displays the order dependence
of both scaled observables. Across integers $2\le N\le15$, the two-to-three-loop
relative change decreases from $1.580\%$ to $0.041\%$ in $N_{f}^\mathrm{cr}$ and from
$1.663\%$ to $0.042\%$ in $a_*$; each is much smaller than the corresponding
one-to-two-loop change. The denominator and weak-coupling scheme are held fixed throughout. The shared one-loop values are exact for the scaled quantities $x$ and $\lambda$; the unscaled $N_f^{\mathrm{cr}}$ and $a_*$ still scale as $N$ and $1/N$. The selected colors $N=2,3,15$ span the finite-$N$ range; the numerical comparison uses every integer $2\le N\le15$.

\begin{table}[t]
\caption{Unfiltered hierarchy at three-loop rational merger flavors ($k=1/2$). Ratios are evaluated at $D=5$; $\Dpt$ uses the first crossing with $\theta=1/2$.}
\label{tab:hierarchy_merger}
\begin{ruledtabular}
\setlength{\tabcolsep}{3pt}\begin{tabular}{cccccc}
$N$ & $N_f$ & $R_2$ & $R_3$ & $R_4$ & $\Dpt$\\
\hline
3 & 6.379443 & 0.163 & 0.269 & 1.012 & 4.440\\
4 & 8.443701 & 0.160 & 0.279 & 0.553 & 4.897\\
5 & 10.519094 & 0.159 & 0.283 & 0.318 & 5.539\\
7 & 14.683750 & 0.158 & 0.287 & 0.098 & 5.738\\
15 & 31.390618 & 0.157 & 0.291 & 0.095 & 5.717\\
\end{tabular}
\end{ruledtabular}
\end{table}

\add{\begin{table}[t]
\caption{Largest dimension $D$ for which the resummation parameter at the four-loop
fixed point satisfies $\xir\le\xir_{\max}$, for $\SU(N)$ with $N_f$ fundamental flavors.
The bounds $\xir_{\max}=0.05,\,0.1,\,0.2$ correspond to enhancements of the one-loop
anomalous dimension by $5\%$, $11\%$ and $25\%$. The weak dependence on $(N,N_f)$
reflects the fact that $\xir$ is controlled primarily by $\epsilon$.}
\label{tab:xi}
\begin{ruledtabular}
\begin{tabular}{ccccc}
$N$ & $N_f$ & $\xir_{\max}=0.05$ & $\xir_{\max}=0.1$ & $\xir_{\max}=0.2$\\
\hline
2 & 0 & 4.12 & 4.26 & 4.59 \\
2 & 2 & 4.10 & 4.21 & 4.46 \\
3 & 0 & 4.12 & 4.26 & 4.59 \\
3 & 2 & 4.11 & 4.22 & 4.50 \\
3 & 4 & 4.09 & 4.19 & 4.41 \\
5 & 5 & 4.10 & 4.21 & 4.45 \\
10 & 10 & 4.10 & 4.21 & 4.45 \\
15 & 20 & 4.09 & 4.19 & 4.40 \\
\end{tabular}
\end{ruledtabular}
\end{table}}

At the next order,
\begin{equation}
 F_4(\lambda,x)=F_3(\lambda,x)-\frac{h_3}{N^4}\lambda^4.
 \label{eq:four_loop_root_polynomial}
\end{equation}
At the $\SU(3)$ three-loop merger flavor, the four-loop matched root is
$a_*^{[4]}=0.0597143$ with
$\partial_a\beta_{[4]}^{(5)}=-0.688650$.
This is a simple UV zero rather than a double zero. 
Figure~\ref{fig:four_loop_uv_persistence} shows how the four-loop correction removes the three-loop merger.

For our convention $\beta=\tfrac12\partial_t a$, a gauge-direction critical
exponent is $\Theta_g=-2\partial_a\beta(a_*)$, since
$\partial_t\delta a=-\Theta_g\delta a$.
Its value is zero at every exact double-root merger by construction.
That zero is not an independent test of convergence. A comparison of
matching orders must hold $(D,N,N_f)$ fixed and follow the same branch,
also checking that the root still exists and remains simple.
For the four-loop root just quoted, $\Theta_g\simeq1.37730$.
The full theory instead requires the eigenvalues of the coupled stability
matrix, so this one-coupling exponent alone does not count all relevant
directions.

Extending Eq.~\eqref{eq:matching_order_family} to five-loop
matching, using the known coefficient $\beta_4$
\cite{herzogFiveloopBetaFunction2017} and keeping the same
denominator $Q(a)=Q_{1/2}(a)$, gives
\begin{align}
 h_4 &= \beta_4-C\beta_3+\frac{C^2}{2}\beta_2,\nonumber\\
 \beta_{[5]}^{(5)}(a)-\beta_{[4]}^{(5)}(a)
     &= -\frac{h_4 a^6}{Q(a)}.
 \label{eq:five_loop_matching_appendix}
\end{align}
At $D=5$, $N=3$, and $N_f=6.379443171$, corresponding to
the three-loop merger benchmark, the five-loop matched flow
has a UV zero at $a_{\mathrm{UV}}^{[5]}\simeq0.0626040$ and
an IR zero at $a_{\mathrm{IR}}^{[5]}\simeq0.157084$.
Using the convention $\Theta_g=-2\partial_a\beta^{(5)}(a_*)$
defined above, the corresponding gauge-direction exponents are
$\Theta_{g,\mathrm{UV}}^{[5]}\simeq1.15342$ and
$\Theta_{g,\mathrm{IR}}^{[5]}\simeq-6.89438$.
Their signs identify UV attraction and IR attraction,
respectively, in the gauge-coupling direction.
An IR partner thus reappears at five-loop matching for the same
$(D,N,N_f)$ at which four-loop matching gives a single UV zero.
This comparison concerns the fixed-denominator rational family;
determining whether the five-loop UV and IR zeros merge requires
tracking both branches as $N_f$ varies.

\begin{table}[t]
\caption{Merger values of $\Nfcr$ and $a_*$ at $D=5$ for $\SU(N)$ with fundamental fermions in the stated three-loop approximations. ``Improved'' refers to
Eq.~\eqref{eq:improved_eta}; the last column keeps the five-dimensional threshold
coefficient $7N-\frac43N_f$ in place of $2\beta_0$ (threshold-input variation, see text).}
\label{tab:main}
\setlength{\tabcolsep}{2pt}
\renewcommand{\arraystretch}{1.1}

\begin{ruledtabular}
\begin{tabular}{cc *{5}{D{.}{.}{1.4}}}

 & & \multicolumn{1}{c}{3-loop}
 & \multicolumn{1}{c}{Improved}
 & \multicolumn{1}{c}{Improved}
 & \multicolumn{1}{c}{Improved}
 & \multicolumn{1}{c}{Threshold} \\

 & $N$
 & \multicolumn{1}{c}{$\MSbar$}
 & \multicolumn{1}{c}{$k^+$}
 & \multicolumn{1}{c}{$k=0.30$}
 & \multicolumn{1}{c}{$k=0.50$}
 & \multicolumn{1}{c}{$k^+$} \\

\hline

$\Nfcr$
 & 2 & 4.845 & 4.308 & 4.314 & 4.345 & 4.124 \\
 & 3 & 7.097 & 6.319 & 6.329 & 6.379 & 6.048 \\

\hline

$a_*$
 & 2 & 0.1639 & 0.1202 & 0.1219 & 0.1301 & 0.1245 \\
 & 3 & 0.1083 & 0.0799 & 0.0810 & 0.0866 & 0.0828 \\

\end{tabular}
\end{ruledtabular}
\end{table}

At fixed $N_f/N$ and $Na$, the three-loop matching equations approach
an $N$-independent planar limit with finite-$N$ corrections.
The baseline merger ratio decreases from $\Nfcr/N\simeq2.17$ at $N=2$
to $\simeq2.09$ at large $N$. Within this three-loop family, varying
$k$ from $(1/4)^+$ to $1/2$ changes $\Nfcr$ by about $1\%$ and
the merger coupling by about $8\%$.

The threshold-input column of Tables~\ref{tab:main} and
\ref{tab:critical_Nf} replaces $2\beta_0=22N/3-4N_f/3$ in the numerator
by $7N-4N_f/3$, holding $b_1$ and $b_2$ fixed.
This lowers the merger estimate by about $4.3\%$ near $k_+$
and $3.7\%$ at $k=1/2$ in the planar limit.
Because the higher numerator coefficients are not rematched,
this is a threshold-input variation, not a consistent higher-loop
scheme conversion. In particular it also changes higher terms when
the full rational function is expanded.
Together with the approximately $10\%$ difference from the unresummed
three-loop boundary, these changes probe selected modeling choices.
They are neither independent errors to be added in quadrature nor a
bound on the next matching-order effect.

\section{Merger tables and auxiliary figures}
\label{app:tables}

Tables~\ref{tab:critical_Nf} and \ref{tab:fixed_point_values} extend Table~\ref{tab:main} to $N\le15$. Each entry solves the positive double-root conditions for its stated beta function. The variation of $\Nfcr/N$ and $Na_*$ resolves finite-$N$ corrections to their limiting values. The threshold column changes only the leading numerator input, as specified in the main text; all other columns use three-loop $\MSbar$ matching. The $k_+$ columns denote limits from the globally pole-free domain $k>1/4$. Fig.~\ref{fig:threshold_sensitivity} shows the dependence of the $\Dpt$ map on the threshold
$\theta$ of Eq.~\eqref{eq:criterion}.

\begin{table}[!t]
\caption{Merger flavor numbers $\Nfcr$ at $D=5$ for $\SU(N)$ with fundamental fermions in the following three-loop approximations: three-loop $\MSbar$ perturbation theory, the improved beta function
\eqref{eq:improved_eta} for three values of the pole parameter $k$, and the
five-dimensional threshold variant of the numerator ($k_+$).}
\label{tab:critical_Nf}
\begin{ruledtabular}
\setlength{\tabcolsep}{3pt}\begin{tabular}{cccccc}
$N$ & 3-loop & $k_+$ & imp.\ $k{=}0.30$ & imp.\ $k{=}0.50$ & thr.\ $k_+$\\
\hline
 2 & 4.845 & 4.308 & 4.314 & 4.345 & 4.124 \\
 3 & 7.097 & 6.319 & 6.329 & 6.379 & 6.048 \\
 4 & 9.385 & 8.361 & 8.374 & 8.444 & 8.002 \\
 5 & 11.688 & 10.415 & 10.431 & 10.519 & 9.967 \\
 6 & 13.997 & 12.474 & 12.494 & 12.600 & 11.938 \\
 7 & 16.310 & 14.537 & 14.559 & 14.684 & 13.911 \\
 8 & 18.626 & 16.601 & 16.627 & 16.769 & 15.887 \\
 9 & 20.942 & 18.667 & 18.696 & 18.856 & 17.863 \\
10 & 23.260 & 20.733 & 20.766 & 20.944 & 19.841 \\
11 & 25.579 & 22.800 & 22.836 & 23.033 & 21.819 \\
12 & 27.898 & 24.868 & 24.907 & 25.122 & 23.798 \\
13 & 30.218 & 26.936 & 26.979 & 27.211 & 25.777 \\
14 & 32.538 & 29.005 & 29.050 & 29.301 & 27.756 \\
15 & 34.859 & 31.073 & 31.122 & 31.391 & 29.735 \\
\end{tabular}
\end{ruledtabular}
\end{table}

\begin{table}[!t]
\caption{Coupling $a_*$ at the annihilation point, same conventions as
Table~\ref{tab:critical_Nf}.}
\label{tab:fixed_point_values}
\begin{ruledtabular}
\setlength{\tabcolsep}{3pt}\begin{tabular}{cccccc}
$N$ & 3-loop & $k_+$ & imp.\ $k=0.30$ & imp.\ $k{=}0.50$ & thr.\ $k_+$\\
\hline
 2 & 0.1639 & 0.1202 & 0.1219 & 0.1301 & 0.1245 \\
 3 & 0.1083 & 0.0799 & 0.0810 & 0.0866 & 0.0828 \\
 4 & 0.0810 & 0.0599 & 0.0607 & 0.0649 & 0.0620 \\
 5 & 0.0647 & 0.0479 & 0.0485 & 0.0519 & 0.0496 \\
 6 & 0.0539 & 0.0399 & 0.0404 & 0.0432 & 0.0413 \\
 7 & 0.0462 & 0.0342 & 0.0347 & 0.0371 & 0.0354 \\
 8 & 0.0404 & 0.0299 & 0.0303 & 0.0324 & 0.0310 \\
 9 & 0.0359 & 0.0266 & 0.0269 & 0.0288 & 0.0275 \\
10 & 0.0323 & 0.0239 & 0.0243 & 0.0259 & 0.0248 \\
11 & 0.0294 & 0.0217 & 0.0220 & 0.0236 & 0.0225 \\
12 & 0.0269 & 0.0199 & 0.0202 & 0.0216 & 0.0206 \\
13 & 0.0248 & 0.0184 & 0.0187 & 0.0199 & 0.0191 \\
14 & 0.0231 & 0.0171 & 0.0173 & 0.0185 & 0.0177 \\
15 & 0.0215 & 0.0159 & 0.0162 & 0.0173 & 0.0165 \\
\end{tabular}
\end{ruledtabular}
\end{table}

\begin{figure*}[tp]
\EMpanel{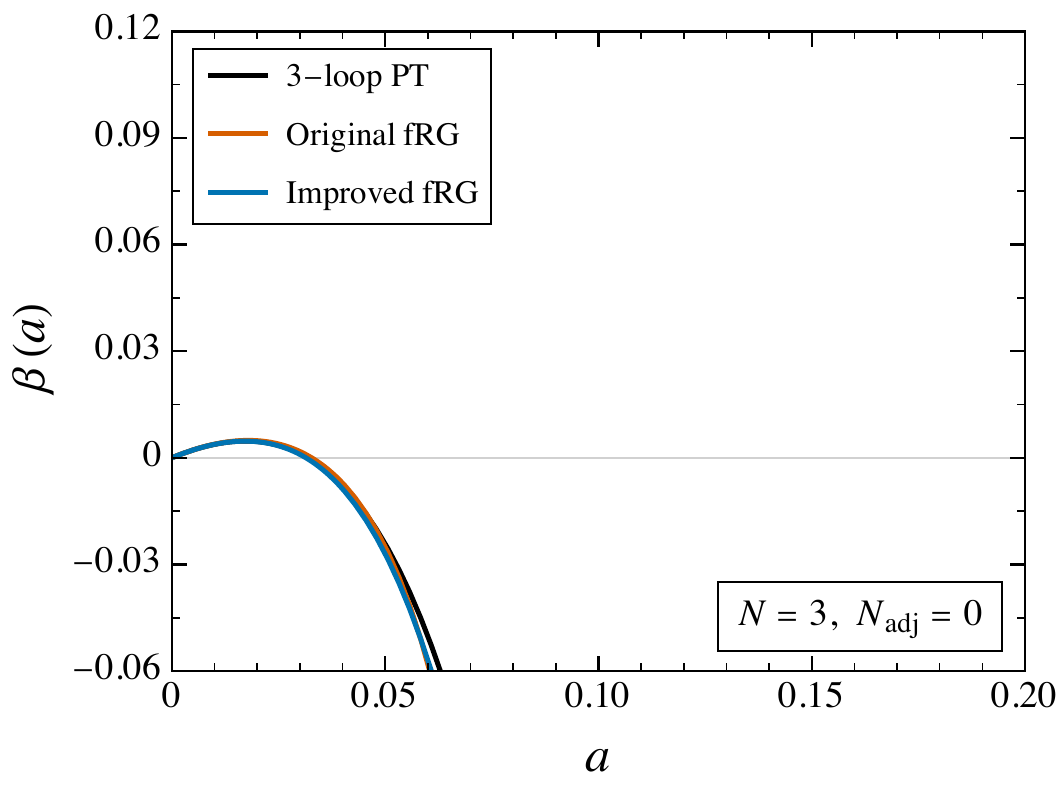}{(a)}{0.2}
\hfill
\EMpanel{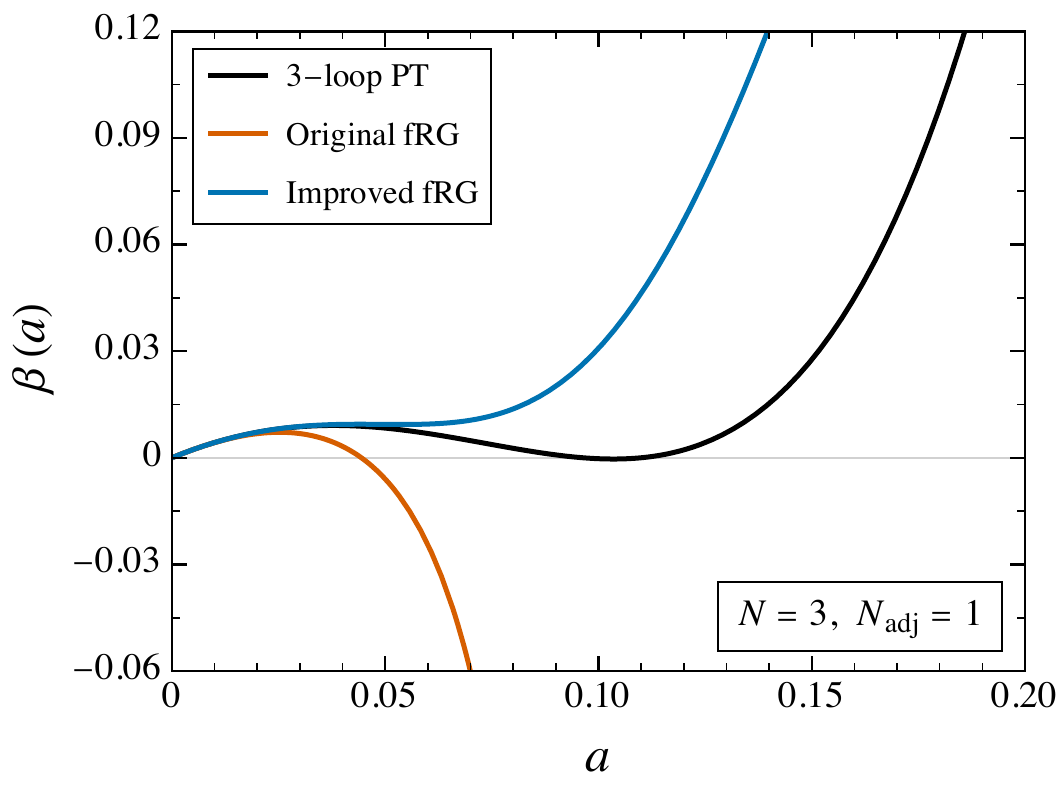}{(b)}{0.2}
\hfill
\EMpanel{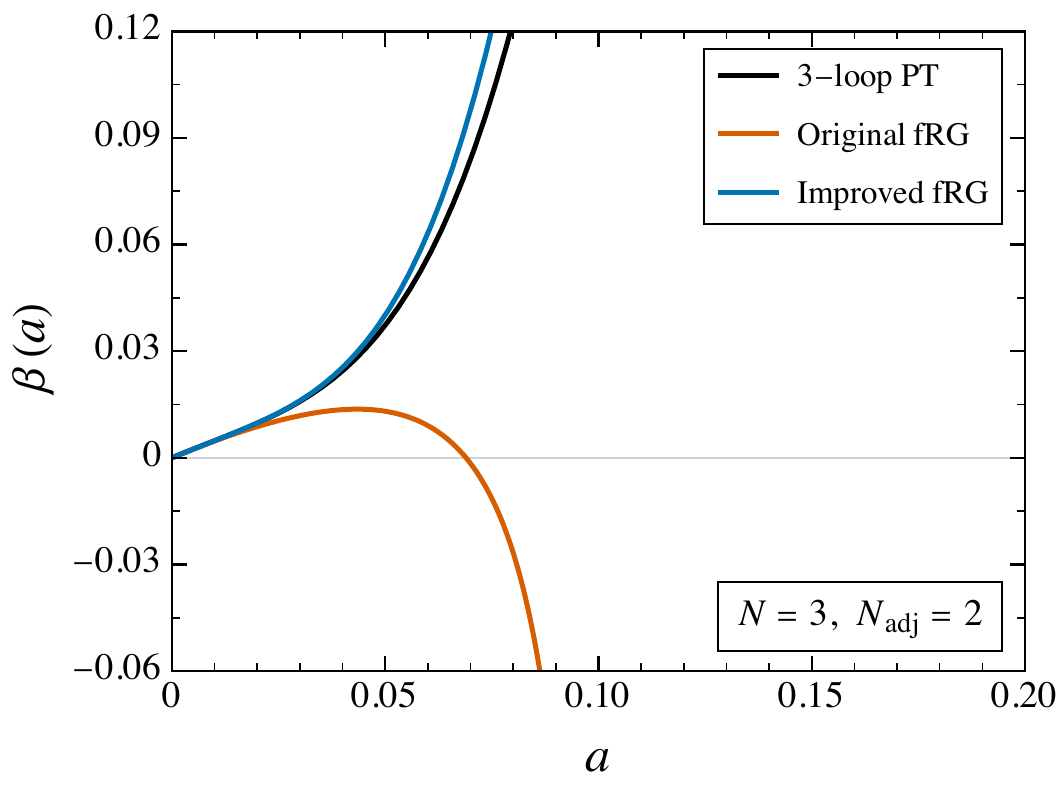}{(c)}{0.2}
\par\vspace{3pt}

% ---------- Two-index symmetric representation ----------
\EMpanel{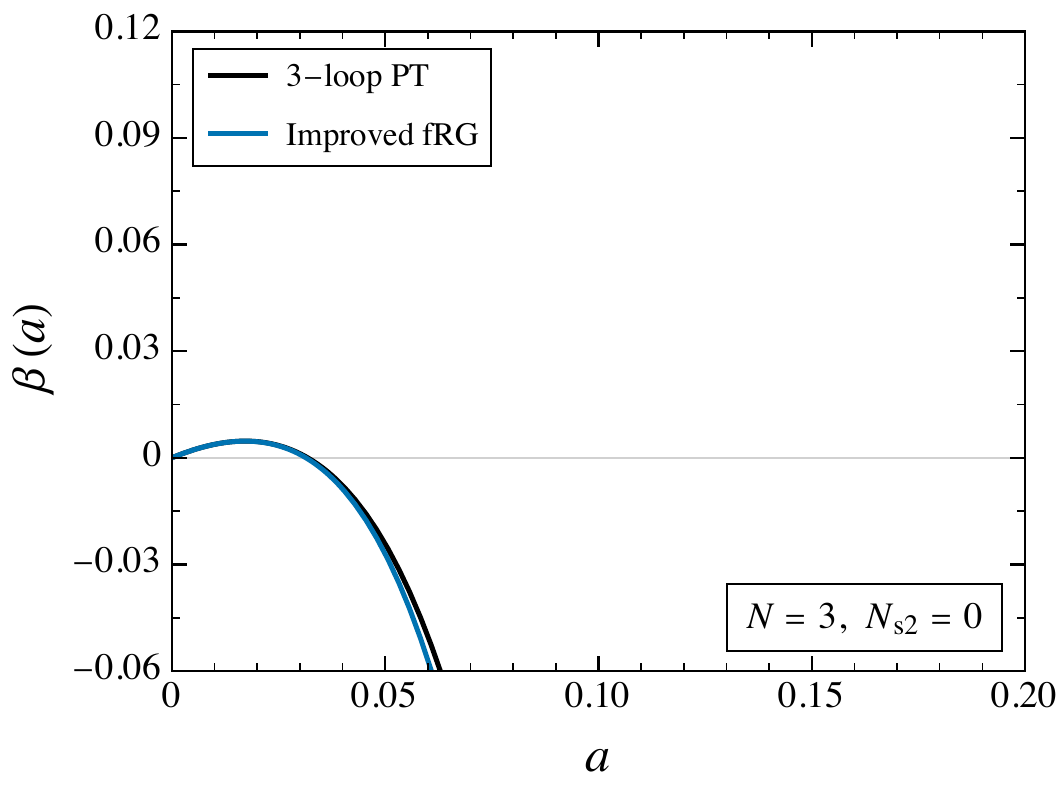}{(d)}{0.2}
\hfill
\EMpanel{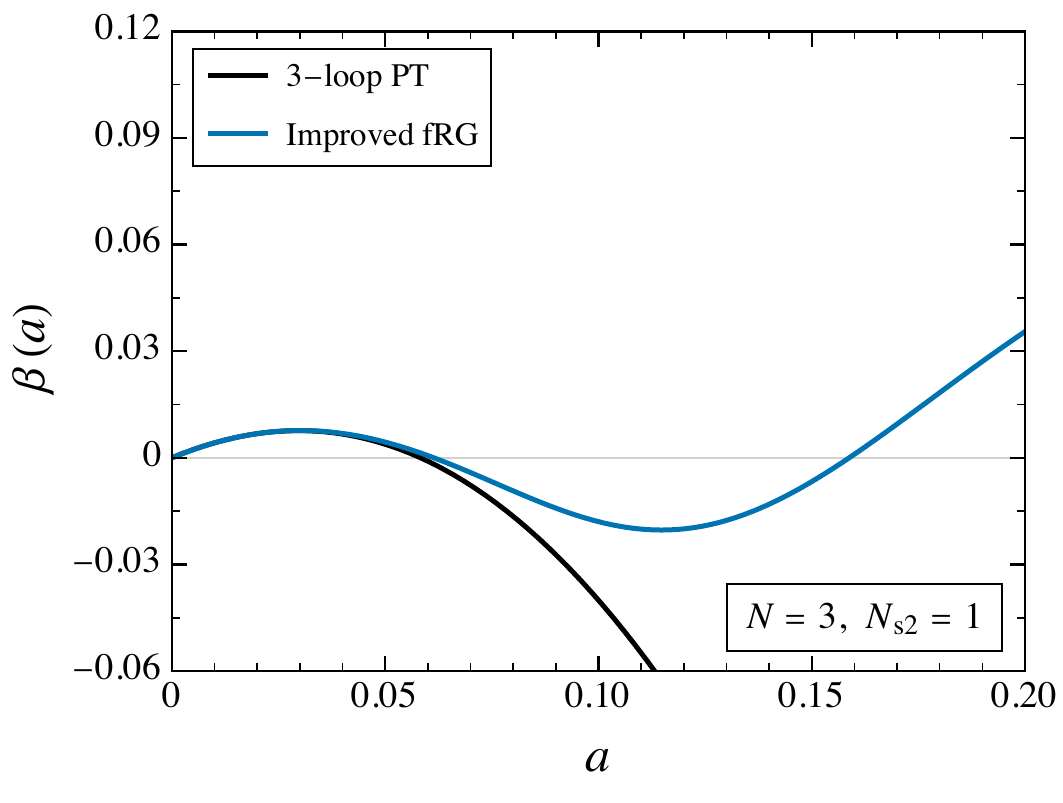}{(e)}{0.2}
\hfill
\EMpanel{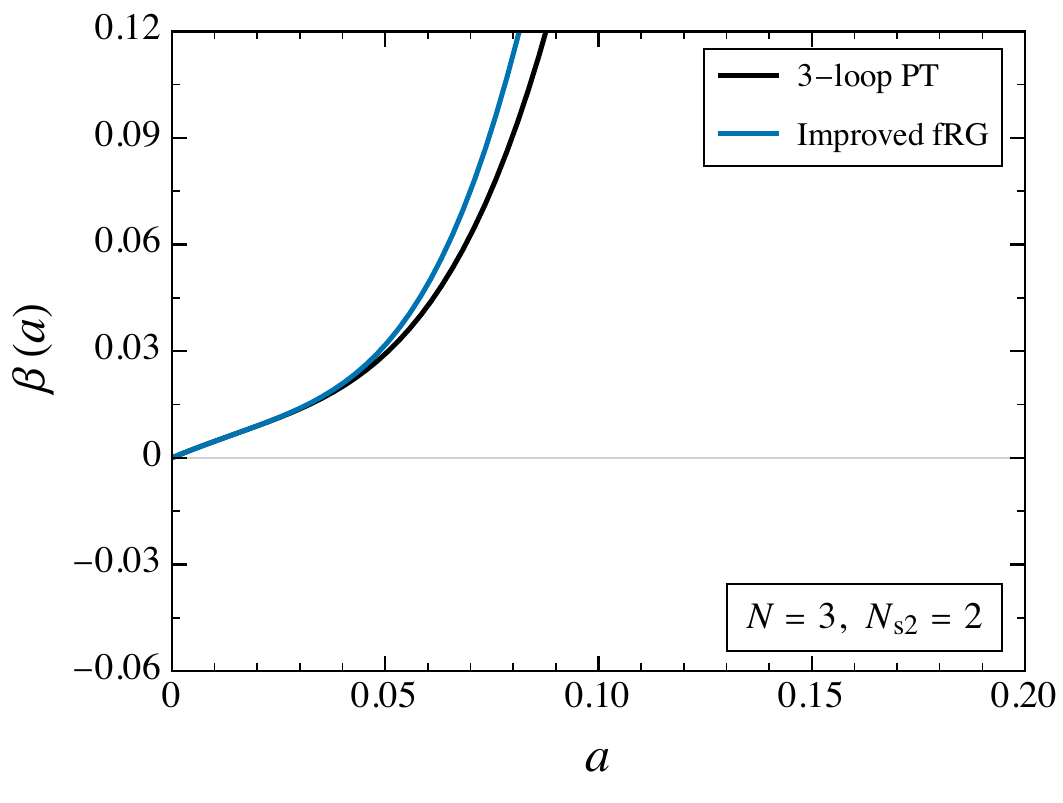}{(f)}{0.2}
\par\vspace{3pt}

% ---------- Two-index antisymmetric representation ----------
\EMpanel{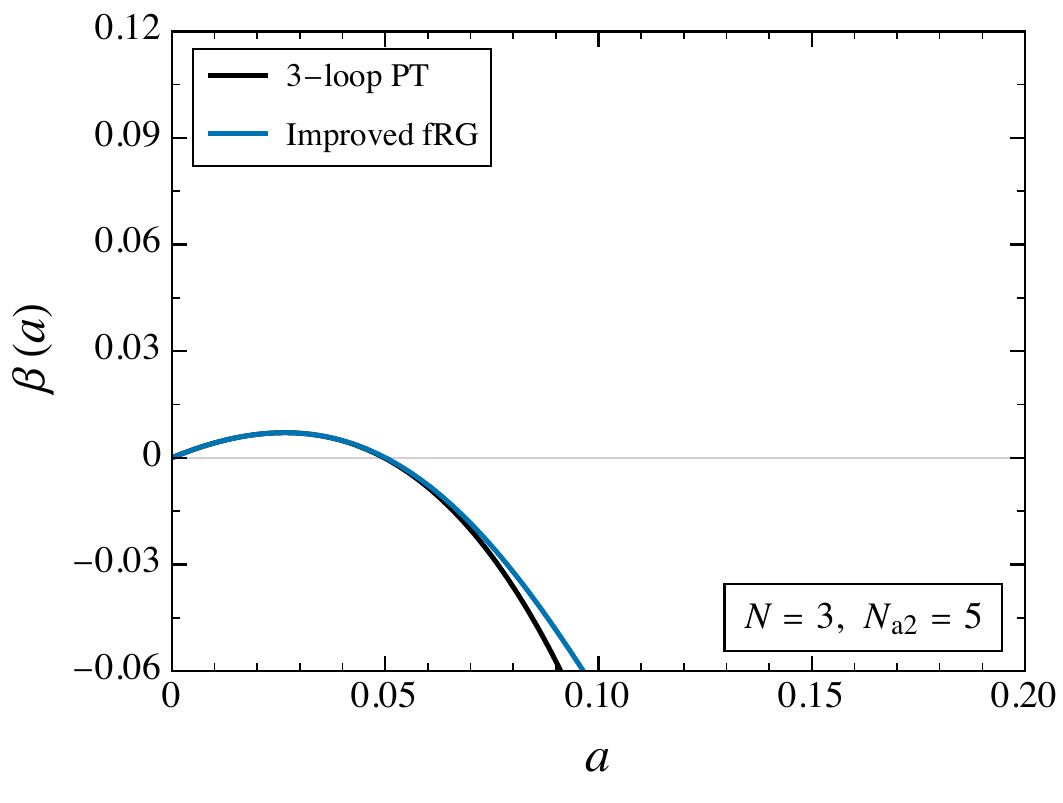}{(g)}{0.2}
\hfill
\EMpanel{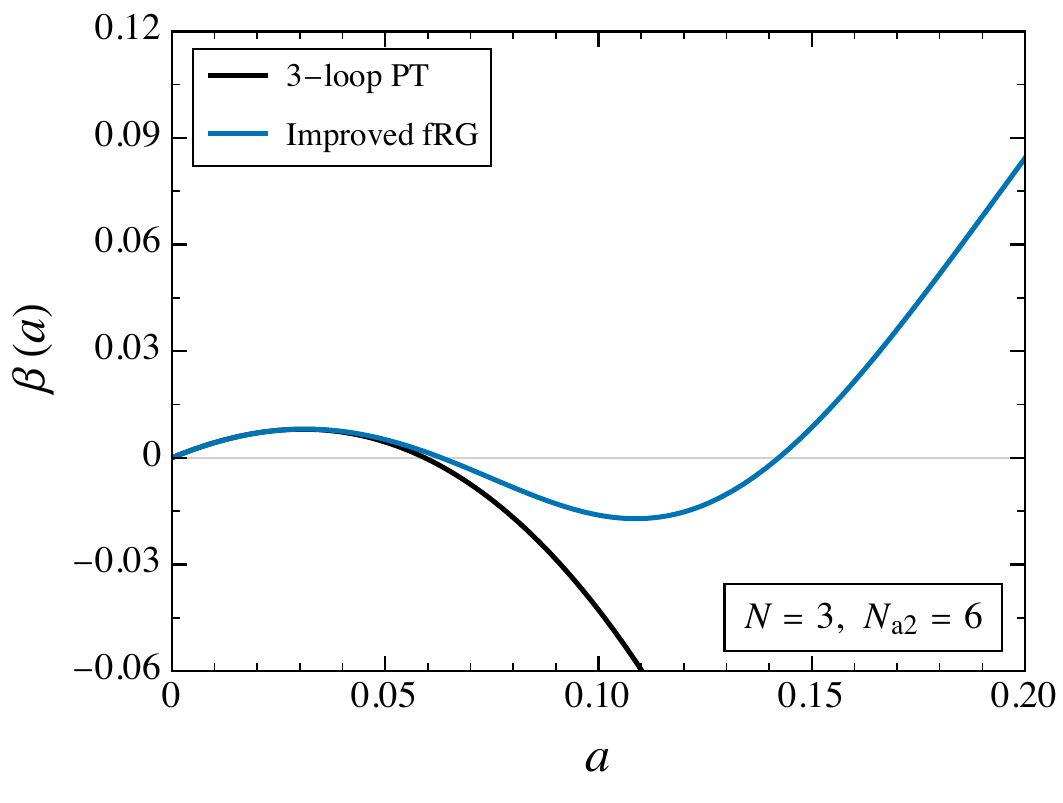}{(h)}{0.2}
\hfill
\EMpanel{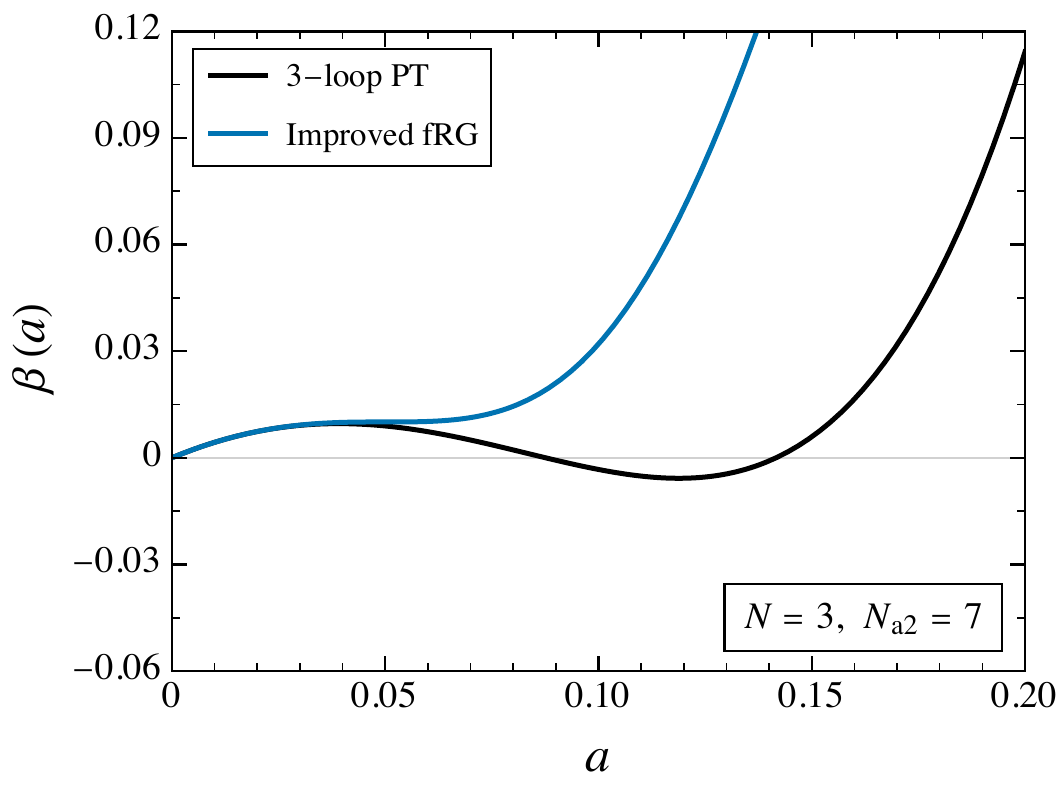}{(i)}{0.2}
\par

\caption{%
Gauge-coupling beta functions at $D=5$ for $\SU(3)$
with fermions in the (a)--(c) adjoint,
(d)--(f) two-index symmetric, and
(g)--(i) two-index antisymmetric representations.
Black and blue curves show the three-loop and improved
results, respectively; orange curves in (a)--(c)
show the leading fRG result.
Fermion multiplicities are indicated in each panel. The rational curves use three-loop matching with $k=1/2$.}
\label{fig:beta_function_representations}

\end{figure*}

Figure~\ref{fig:threshold_sensitivity} compares the hierarchy maps for
three choices of $\theta$. Raising the threshold relaxes the diagnostic
and brings the main $\Dpt=5$ contour closer to the estimated merger
curves. Figure~\ref{fig:beta_function_representations} illustrates the
representation dependence of the gauge-coupling roots at fixed $N=3$.
The two-index antisymmetric representation is conjugate to the
fundamental for $\SU(3)$ and provides a consistency check.
These illustrative beta-function plots can not be used to extend the fundamental hierarchy scan to other representations.

\end{document}